\documentclass[prd,aps,floatfix,twocolumn,superscriptaddress,preprintnumbers]{revtex4-2}

\usepackage[paperwidth=22cm,paperheight=28.7cm,left=1.5cm,right=1.5cm, top=2.cm,bottom=2.cm]{geometry}

\usepackage{amsmath,mathtools,cases,slashed,bm}
\usepackage[colorlinks=true,citecolor=blue,linkcolor=blue,urlcolor=blue]{hyperref}
\usepackage[dvipsnames]{xcolor}
\usepackage{tabularx}
\usepackage{hhline}
\usepackage[normalem]{ulem} 
\usepackage{lipsum}

\usepackage{graphicx}
\usepackage{subfigure}
\usepackage{epsfig}
\usepackage{dcolumn}
\usepackage{bm}
\usepackage{array}
\usepackage{slashed}
\usepackage{braket}
\usepackage{multirow}
\usepackage{bigstrut}
\usepackage{caption}
\usepackage{bbding}
\usepackage{textgreek}
\usepackage{wasysym}
\usepackage{latexsym}
\usepackage{stmaryrd}
\usepackage{txfonts}
\usepackage[T1]{fontenc}
\usepackage{textcomp}
\usepackage{booktabs}
\usepackage{titlesec}
\usepackage{comment}
\usepackage{color}
\usepackage{float}
\usepackage{setspace}
\usepackage{enumerate}

\graphicspath{{./figure/}}
\definecolor{lightgray}{gray}{0.85}

\newcommand{\ubar}{\bar{u}}

\newcommand{\dbar}{\bar{d}}

\newcommand{\sbar}{\bar{s}}
\newcommand{\cbar}{\bar{c}}
\newcommand{\tc}{\textcolor}

\def\hera{HERA I+II $e^{\pm}p$}
\def\atlaspdf{ATLAS-epWZ16}
\def\herapdf{HERAPDF2.0}

\def\xfitter{\texttt{{\small X}F{\small ITTER}}}
\def\qcdnum{\texttt{QCDNUM}}
\def\mcfm{\texttt{MCFM}}
\def\applgrid{\texttt{APPLGRID}}
\def\minuit{\texttt{MINUIT}}

\def\fitone{\texttt{Fit1}}
\def\fittwo{\texttt{Fit2}}
\def\fitonep{\texttt{Fit1p}}
\def\pkushape{\texttt{PKU-fit}}

\begin{document}

\newcommand*{\hpdf}{HERAPDF2.0}

\author{Alim~Ruzi}
\affiliation{\PKU}
\author{Bo-Qiang Ma}
\email{mabq@pku.edu.cn} \affiliation{\PKU}\affiliation{\CHEP}\affiliation{\CICQM}

\newcommand*{\PKU}{School of Physics and State Key Laboratory of Nuclear Physics and Technology, Peking University, Beijing 100871, China}
\newcommand*{\CHEP}{Center for High Energy Physics, Peking University, Beijing 100871, China}
\newcommand*{\CICQM}{Collaborative Innovation Center of Quantum Matter, Beijing, China}

\title{Reinterpretation of proton strangeness between ATLAS and CMS measurements}\thanks{Phys. Rev. D 104 (2021) 076004, \url{https://link.aps.org/doi/10.1103/PhysRevD.104.076004}}
\begin{abstract}
We reexamine the shapes of the strange quark parton distribution functions (PDFs) of the proton by means of QCD analysis of {\hera} deep inelastic scattering cross section measurement at DESY, and inclusive gauge boson production and $W$ boson production associated with a charm quark from the LHC at CERN. We find that there is an overall agreement on the strange quark distributions obtained from CMS $W$ + charm and ATLAS $W/Z$ data at the parton momentum fraction range $x \lesssim 10^{-2}$. Meanwhile, there is also a strong tension between these data toward large $x$.  We find that this tension fades away if the ATLAS measurement of $W/Z$ production is analyzed together with the ATLAS $W$ + charm data. The $W/Z$ and $W$ + charm data both from ATLAS and CMS experiments agree that the proton strangeness is enhanced toward small momentum fraction $x$ and is smoothly suppressed at large $x$. Furthermore, a strong $x$ dependence of the strange-to-nonstrange parton ratio $R_s(x,Q^2)$ is observed.
\end{abstract}


\maketitle

\section{Introduction}

It is well known that nucleons consist of pointlike particles denominated as partons, revealed by lepton nucleon deep inelastic scattering (DIS) experiment~\cite{Bloom:1969kc,Breidenbach:1969kd} almost half a century ago. There have been tremendous efforts to precisely determine the parton distribution functions (PDFs) of the proton worldwide. Distributions of the valence-up quarks, valence-down quarks and gluons are determined to a level that could give, by and large, a consistent result on physical quantities in the standard model (SM) obtained from high-energy collision experiments~\cite{DelDebbio:2007ee, Dulat:2015mca, Harland-Lang:2014zoa, Alekhin:2009ni, CooperSarkar:2011aa}. However, the strange content of the proton, which is poorly known because of limited sources of strangeness-sensitive data, is still ambiguous to our understandings of the partonic structure of the proton. Strange quark PDFs of nucleons play a vital role for a number of physics processes at the LHC, ranging from measurements of $W$ boson production in association with a charm jet~\cite{Baur:1993zd} and determinations of electroweak (EW) interaction parameters to the formation of strange hadrons~\cite{Farhi:1984qu}.

The size and shape of the strange quark PDFs have recently attracted a lot of interest and ignited a hot debate on specialized strange studies. The strange quark PDFs obtained from dimuon production in neutrino-nucleus DIS experiment by the NuTeV and NOMAD Collaborations~\cite{Goncharov:2001qe, Mason:2007zz, Samoylov:2013xoa}, are significantly suppressed while the inclusive $W/Z$ production data by the ATLAS Collaboration~\cite{Aad:2012sb} give strongly enhanced distribution of strangeness. The ratio of strange to nonstrange sea quark PDFs, $R_s$ = $(s+\sbar)/(\ubar + \dbar)$, from neutrino-nucleus DIS experiment is approximately 0.5 when evaluated at $x$ = 0.023 and energy scale $Q$ = 1.6~GeV, whereas the strange to antidown quark ratio, $r_s = (s+\sbar)/2\dbar$, evaluated at $x$ = 0.023 and $Q^2$ = 1.9 $\mathrm{GeV^2}$ from the ATLAS analysis of the inclusive gauge boson production together with the combined {\hera} data~\cite{Abramowicz:2015mha}, rises up to unity.

In their QCD analysis of accumulated $W/Z$ data to extract {\atlaspdf}~\cite{Aaboud:2016btc}, the ATLAS Collaboration obtains even larger strange-to-nonstrange ratio, $R_s$ = 1.131~\cite{Aaboud:2016btc}, showing agreement with their older analysis~\cite{Aad:2012sb}. Another study of the ATLAS and CMS $W/Z$ cross section measurements in Ref.~\cite{Cooper-Sarkar:2018ufj} reports that there is no tension between the LHC $W/Z$ and the HERA data or between the ATLAS and CMS datasets, indicating that the LHC measurements of inclusive $W/Z$ boson production support unsuppressed strangeness in the proton at low $x$ at both low-and high-energy scales.

A recent global analysis of vector gauge boson production in association with jets plus older $W/Z$ and HERA $e^{\pm}$ data by the ATLAS Collaboration~\cite{ATLAS:2021qnl} reports somewhat different $R_s$ distribution from the one in {\atlaspdf} PDF sets. The ratio $R_s$ in~\cite{ATLAS:2021qnl} keeps consistency with {\atlaspdf} at $x \lesssim 10^{-2}$ and declines faster at other $x$ value. But, the strange quark in that analysis still stays comparable or even larger than the nonstrange sea quarks. The very recent global PDFs, MHST20 PDFs~\cite{Bailey:2020ooq}, extracted from almost all of the available measurements show an $R_s$ value compatible with that of the {\atlaspdf}.

So it comes as a puzzle that both the ATLAS and CMS measurements of $W/Z$ boson production support an enhanced strangeness inside the proton, while the fixed target lepton-nuclear collision prefers a rather suppressed strangeness. Among all the standard model processes, the $W s$ $\to$ $c$  interactions have the utmost sensitivity to the strange (antistrange) quark PDFs inside the proton. The QCD analysis of $W$ production in association with a charm (anticharm) quark in the proton-proton  (pp) collision at $\sqrt{s}$ = 7~TeV by the ATLAS Collaboration shows the ratio of strange to down sea quark, $r_s = (s+\sbar)/2\dbar$, to be 0.96~\cite{Aad:2014xca}, supporting the results obtained from the analysis of $W/Z$ measurements. However, the CMS Collaboration, in the analysis of the same process, $W$  + charm production plus $W$-lepton charge asymmetry~\cite{Chatrchyan:2013mza, Khachatryan:2016pev, Chatrchyan:2013uja,Sirunyan:2018hde}, reports that they have not observed the enhanced strangeness as the ATLAS Collaboration did. Instead, they observed an $x$-dependent and suppressed ratio $R_s$. Yet, in another dedicated analysis on the strange quark distribution by the Hessian error updating method~\cite{Yalkun:2019gah,Schmidt:2018hvu}, it is reported that the CMS $W$ + charm quark production data~\cite{Chatrchyan:2013uja} enhance strangeness of CT14 PDFs families. The recently extracted CT18 PDFs series from fixed target and up-to-date collider data by the CTEQ group~\cite{Hou:2019efy} demonstrate that the strange quark density is enhanced over the nominal fit CT18NNLO when the inclusive $W/Z$ data are included. Another devoted study on the strangeness of the proton~\cite{Faura:2020oom}, which includes both the lepton-nucleus DIS cross section data from the NuTeV and NOMAD plus $W$ + charm data from the ATLAS and CMS experiments, reports a slightly enhanced strange quark PDF within uncertainties in terms of the ratio $R_s(x, Q^2)$ with huge uncertainties at large momentum fraction.

Given the important influence of the strange quark PDFs in SM physics inside the LHC, and the contradictory interpretations of the strange-sensitive data from the ATLAS and CMS measurements, we regard reinvestigation of the exact shapes of the strange quark PDFs of the nucleon by a global analysis of strangeness-sensitive measurements, within an accurate theoretical and methodological framework, as necessary to clarify the controversial interpretations of the strangeness-related quantities, such as $xs(x, Q^2)$ and the ratio $R_s$. This paper is intended to fulfill this purpose. We present a fairly detailed PDFs extraction with the general-purpose PDFs fitting  program {\xfitter}~\cite{Alekhin:2014irh} from the following data: HERA I+II $e^{\pm}p$~\cite{Abramowicz:2015mha}, ATLAS combined inclusive $W/Z$ production cross section~\cite{Aad:2012sb,Aaboud:2016btc}, and $W$ production in association with a charm quark~\cite{Aad:2014xca}; CMS $W$-lepton charge asymmetry~\cite{Chatrchyan:2013mza, Khachatryan:2016pev} and $W$ production in association with charm quark~\cite{Chatrchyan:2013uja,Sirunyan:2018hde}, applying variants of different parametrization forms first to clarify whether the same physics processes in ATLAS and CMS experiments give compatible or incompatible distributions of the strange quark and, second, to pin down its exact shape.

\section{Input data and corresponding theoretical predictions}\label{sec:sec2}
\subsection{Data description}\label{subsec:data}
The process of $e^{\pm}p$ DIS has been the central dataset to many PDFs working groups in their global analyses and provides the most constraining power to quark PDFs when parametrized as a composite functional form. The gauge boson production data at the LHC also provide significant constraints on the flavor decomposition where the HERA data fail. The $W$ + charm production offers the best constraints on the strange quark PDFs and gluon distributions through the subprocesses: $g+s$ $\to$ $W^-$ + $c$ and $\sbar + g$ $\to$ $W^+$ + $\cbar$. The $W$-lepton charge asymmetry data from CMS, available for $7~$ and $8~\mathrm{TeV}$, can constrain the valence quarks very well and have an indirect impact on the strange content.

We do not include neutrino-induced dimuon data in our analysis. The dimuon data were obtained from $\nu_{\mu}(\bar{\nu}_{\mu})A$ fixed target DIS process where the target is iron, whose internal structure is more complex than the free nucleon. In this process, however, the nuclear European Muon Collaboration (EMC) effect should be taken into account in addition to some nuclear corrections; see Ref.~\cite{Gong:2018mzx} and references therein. Considering the non-negligible nuclear corrections in the lepton-nuclear collision experiment, we believe that pure-proton-related data, which are free of the EMC effect, are significantly useful for our analysis. Besides that, the strange quark distribution extracted from dimuon production in neutrino scattering differs significantly from the one extracted from LHC data, especially the ATLAS $W/Z$ production process. The latter prefers a rather higher $R_s$ value and the former prefers a lower value. One more disadvantage of dimuon data is that their constraining power vanishes at $x \lesssim 0.01$ due to shadowing and anti-shadowing effects in neutrino-nucleus DIS process~\cite{Brodsky:2004qa}.

The description of data is given below:
\begin{enumerate}[(i)]
\item The {\hera} combined data~\cite{Abramowicz:2015mha} include measurements acquired from run 1 during 1992 to 2000, and run 2 during 2002 to 2007  in $e^{\pm}p$ collision at different electron/positron and proton beam energies with approximately the total integrated luminosities of 500 pb$^{-1}$.  The datasets are divided between $e^{\pm}p$ charged current (CC) and neutral current (NC) scattering according to the intermediate particle being $W^{\pm}$ or $Z/\gamma^*$. The $e^{\pm}p$ CC and $e^-p$ NC interaction scattering cross sections are measured at $\sqrt{s}$ = 318~GeV. The rest of the datasets are made up of $e^+p$ NC scattering cross sections with positron
beam energy of $\mathrm{E_e}$ = 27.5~GeV and proton beam energies of $\mathrm{E_p}$ = 920, 820, 575 and 460~GeV, corresponding to total center-mass energies $\sqrt{s}$ = 318, 300, 251, and 225~GeV. These datasets when combined give in total 1307 data points on differential cross sections being a function of the Bjorken variable $x$, approximately the light-cone momentum fraction of the quark, and the transferred-energy squared $Q^2$. For the published datasets, these two kinematic variables can cover $6 \cdot 10^{-7} \leq x \leq 0.65$, $ 0.045 \leq Q^2 \leq 50000$ $\mathrm{GeV^2}$ in the NC interaction scattering and $1.3 \cdot 10^{-2} \leq x \leq 0.40$, $200 \leq Q^2 \leq 50000$ $\mathrm{GeV^2}$ in the CC interaction scattering.
	
	\item The ATLAS $W/Z$ data~\cite{Aaboud:2016btc} include the cross section measurements of $W^{\pm}$ production in the leptonic decay channels: $W^+\to  l^+\nu$,~ $W^-\to l^-\bar{\nu}$ and $Z \to ll$ ($l=e, \mu$) production processes at $\sqrt s =$ 7~TeV with an integrated luminosity of $4.6$~fb$^{-1}$. The differential cross sections for $W^{\pm }$ are given as the function of lepton pseudorapidity in the range $|\eta_l|< $ 2.5 with 22 data points in total. The cross section of $Z/\gamma^*$ production is measured as a function of the absolute dilepton rapidity $|y_{ll}|$ in the central and forward region, for three intervals of dilepton invariant mass or the intermediate $Z$ boson mass, $ 46 < m_{ll} < 66$~GeV, $66 < m_{ll} < 116$~GeV (forward and central rapidity), and $ 116 < m_{ll} < 150$~GeV (forward and central rapidity). The $Z/\gamma^*$ data include 39 data points in total.
	\item The data of $W$ + charm measurement come from ATLAS and CMS experiments. The ATLAS measurements of $W^{\pm}$ production in association with an (anti)charm quark are available  for $\sqrt{s}$ = 7~TeV with total 22 data points~\cite{Aad:2014xca}, while the CMS data of this process are available for 7~\cite{Chatrchyan:2013uja} and 13~TeV~\cite{Sirunyan:2018hde} averaged for $W^+$ + $\cbar$ and $W^-$ + c with a total of five  points. Both ATLAS and CMS measurements of $W$ + charm production are given as function of lepton pseudorapidity, $\eta^{\mu}$.
	\item The $W$-lepton charge asymmetry, defined as $\mathcal{A}$ = $\frac{\sigma(W^+) - \sigma(W^-)}{\sigma(W^+) + \sigma(W^-)}$, from CMS is available for 7~\cite{Chatrchyan:2013mza} and 8~TeV~\cite{Khachatryan:2016pev}.
\end{enumerate}

\subsection{Theoretical calculation}
The corresponding predictions of $e^{\pm}p$ cross sections for {\hera} data~\cite{Abramowicz:2015mha} are obtained by solving the DGLAP evolution equations~\cite{Gribov:1972ri,Gribov:1972rt,Dokshitzer:1977sg,Altarelli:1977zs} at next-to-next-to-leading-order (NNLO) in the $\overline{\mathrm{MS}}$ scheme~\cite{Fanchiotti:1992tu} through the \qcdnum~program~\cite{Botje:2010ay} interfaced with \xfitter. The renormalization and factorization scales are chosen to be $\mu_r^2$ = $\mu_f^2$ = $Q^2$. The initial factorization scale is set to be $Q_0^2 = 1.9~\mathrm{GeV}^2$ just below the charm quark mass $m_c = 1.43~\mathrm{GeV}$ in this analysis. The evolved PDFs at each scale point are convoluted with the coefficient functions of the structure functions of the proton to obtain the corresponding differential cross sections for $e^{\pm}p$ scattering. The contributions to the structure functions from heavy quarks are calculated in the general-mass-variable-flavor-number scheme~\cite{Thorne:1997ga,Thorne:2006qt,Thorne:2012az} in the NC interactions. For the structure functions in CC interactions, the zero-mass approximation is used because the HERA CC data have $Q^2\gg M_H^2$, where $M_H$ is the heavy quark mass, i.e., the masses of charm and bottom quarks. The calculation of the differential cross section for the gauge boson production is done using the well-known Monte Carlo program \mcfm~\cite{Campbell:1999ah,Campbell:2011bn} with the CT18NNLO PDF set~\cite{Hou:2019efy}. The $W^{\pm}$ and $Z$ boson cross sections are available up to NNLO. Because of the intensiveness of this calculation, a fast convolution technique is applied. First of all, the partonic cross sections in $pp$ collision are obtained in the form of grid files using {\mcfm} with the help of another fast parton convolution program {\applgrid}\cite{Carli:2010rw} interfaced both with {\mcfm} and {\xfitter}; then, the corresponding cross sections are obtained convoluting the grid files with the evolved PDFs during the fit. So far, {\applgrid} (version-1.5.34) can only calculate cross sections up to next-to-leading-order (NLO) accuracy. To get the NNLO total and differential cross sections, a $K$-factor formalism is applied. The $K$-factor is defined as
\begin{equation}
	K= \frac{\sigma^{\mathrm{NLO~EW}}_{\mathrm{NNLO~QCD}}} {\sigma^{\mathrm{LO~EW}}_{\mathrm{NLO~QCD}}}
\end{equation}
For both the CMS and ATLAS $W$ + charm data, only the NLO calculations are used and this is obtained via {\mcfm}.
The theoretical prediction for $W$ + charm process is only available at NLO calculation, while that for the Drell-Yan (DY) and $W^{\pm}$ production can be obtained up to NNLO.  The measurements for the $W/Z$ data~\cite{Aaboud:2016btc} are given in high precision, so it is important to get as accurate theoretical predictions as possible. During the fit, the order of calculation is specified according to the physics process under consideration.
All the predictions are calculated in the respective fiducial phase space of the experimental data. The $K$-factors are evaluated bin by bin with the same PDFs in both the numerator and denominator.

\section{Analysis setup}

We are curious about the ratio $R_s$ given by CMS analysis of $W$ + charm production together with $W$-lepton charge asymmetry data~\cite{Sirunyan:2018hde} being much different from the one reported by the ATLAS group~\cite{Aaboud:2016btc}. The CMS analysis~\cite{Sirunyan:2018hde} made a comparison of the ratio $R_s$ to the one obtained with {\atlaspdf} PDFs~\cite{Aaboud:2016btc} and ABMP16 PDFs~\cite{Alekhin:2018pai}. $R_s$ from ATLAS is almost constant throughout the entire $x$ except at high $x$ and the property of being greater than 1 is interpreted as enhancement of strange quark PDFs relative to nonstrange quark PDFs by the ATLAS group. On the contrary, $R_s$ of CMS is fully $x$ dependent and declining faster as $x$ getting larger, interpreted as suppressed strangeness by CMS. This is the main difference between ATLAS $W/Z$ and CMS $W$ + charm analyses in terms of the ratio $R_s$. Considered as dominant data sources in constraining the light flavor quark and antiquark PDFs, the ATLAS $W/Z$ and CMS $W$ + charm data should not be controversial as far as the strange quark PDFs are concerned. Below, we are going to explore both the analysis setup of the ATLAS~\cite{Aaboud:2016btc} and CMS~\cite{Sirunyan:2018hde} analyses.
The {\atlaspdf}  is extracted at NNLO  accuracy in perturbative QCD using inclusive {\hera}~\cite{Abramowicz:2015mha} and ATLAS $W/Z$ data~\cite{Aaboud:2016btc} through {\xfitter} framework. The PDFs in {\atlaspdf} are parametrized at a starting scale $Q^2_0$ = 1.9~$\mathrm{GeV}^2$  as
\begin{subequations}
	\begin{align}
		xg(x)   & = A_gx^{B_g}(1-x)^{C_g}-A'_gx^{B'_g}(1-x)^{C'_g}, \label{eq:xg}  \\
		xu_v(x) & = A_{u_v}x^{B_{u_v}}(1-x)^{C_{u_v}}(1+E_{u_v}x^2),\label{xq:xuv}\\
		xd_v(x) & = A_{d_v}x^{B_{d_v}}(1-x)^{C_{d_v}},              \label{eq:xdv}\\
		x\ubar(x) & = A_{\ubar}x^{B_{\ubar}}(1-x)^{C_{\ubar}},      \label{eq:xbaru}\\
		x\dbar(x) & = A_{\dbar}x^{B_{\dbar}}(1-x)^{C_{\dbar}},\label{eq:xbard} \\
		x\sbar(x) & = xs(x) = A_{\sbar}x^{B_{\sbar}}(1-x)^{C_{\sbar}}.\label{eq:xbars}
	\end{align}
	\label{eq:atasparam}
\end{subequations}
The relationship $A_{\ubar}$ = $A_{\dbar}$ and $B_{\ubar}$ = $B_{\dbar}$ = $B_{\sbar}$ are set for the sea quark parameters based on the assumption $\ubar$ = $\dbar$ as $x\to0$, complying with the usual way quoted by the HERA Collaboration~\cite{Abramowicz:2015mha}.

On the contrary, the CMS Collaboration applies similar shape functions at $Q^2_0$ = 1.9 $\mathrm{GeV^2}$ for their analysis of $W$ + charm quark production together with the $W$-lepton charge asymmetry data to extract PDFs at NLO using {\xfitter}. The applied parametrization form in the CMS analysis is:
\begin{subequations}
	\begin{align}
		xg(x)   & = A_gx^{B_g}(1-x)^{C_g}, \label{eq:xg}  \\
		xu_v(x) & = A_{u_v}x^{B_{u_v}}(1-x)^{C_{u_v}}(1+E_{u_v}x^2),\label{xq:xuv}\\
		xd_v(x) & = A_{d_v}x^{B_{d_v}}(1-x)^{C_{d_v}},\label{eq:xdv}\\
		x\ubar(x) & = A_{\ubar}x^{B_{\ubar}}(1-x)^{C_{\ubar}}(1+E_{\ubar}x^2), \label{eq:xbaru}\\
		x\dbar(x) & = A_{\dbar}x^{B_{\dbar}}(1-x)^{C_{\dbar}},\label{eq:xbard} \\
		x\sbar(x) & = xs(x) = A_{\sbar}x^{B_{\sbar}}(1-x)^{C_{\sbar}}.\label{eq:xbars}
	\end{align}
	\label{eq:cmsparam}
\end{subequations}
In their parametrization form, the CMS Collaboration applies no extra relationship between sea quark parameters except for the renormalization parameters of $\ubar$ and $\dbar$, $A_{\ubar} = A_{\dbar}$. They put an extra parameter that directly affects the strange quark PDFs: strangeness fraction $f_s = \sbar/(\dbar + \sbar)$, often used by the HERA group in their PDFs analysis~\cite{Abramowicz:2015mha} as well. This strangeness fraction number is set as a free parameter in the fit. Power parameters $B$ of all the light flavor sea quarks, $\ubar$, $\dbar$, $\sbar$ are also set free from each other.

To know  whether there is a tension between ATLAS $W/Z$ and CMS $W$ + charm data on the strange-related distributions, we conduct series of fits to the {\hera}, ATLAS and  CMS $W$ + charm data together with the available $W$-lepton charge asymmetry measurement, applying both Eqs.~(\ref{eq:atasparam}) and (\ref{eq:cmsparam}) using the {\xfitter} program. After clarifying the reason behind the contradictory results between ATLAS and CMS data, it is necessary to pin down what the distribution of strange quark PDFs looks like. To achieve this, we consider that the combined usage of all the data available here in a fit with a flexible parametrization form to accommodate the experimental uncertainties is much preferable to the analysis of merely ATLAS or CMS data using Eqs.~(\ref{eq:atasparam}) and~(\ref{eq:cmsparam}). In this way, the strange quark PDFs could be even more constrained. Thus, we do one more round of fit with the following parametrization form at $Q_0^2 = 1.9~{\mathrm{GeV^2}}$ including all the data at once
\begin{subequations}
	\begin{align}
		xg(x)   & = A_gx^{B_g}(1-x)^{C_g}-A'_gx^{B'_g}(1-x)^{C'_g}, \label{eq:xg}  \\
		xu_v(x) & = A_{u_v}x^{B_{u_v}}(1-x)^{C_{u_v}}(1+E_{u_v}x^2),\label{xq:xuv}\\
		xd_v(x) & = A_{d_v}x^{B_{d_v}}(1-x)^{C_{d_v}},              \label{eq:xdv}\\
		x\ubar(x) & = A_{\ubar}x^{B_{\ubar}}(1-x)^{C_{\ubar}}(1+D_{\ubar}x),      \label{eq:xbaru}\\
		x\dbar(x) & = A_{\dbar}x^{B_{\dbar}}(1-x)^{C_{\dbar}}(1+D_{\dbar}x),\label{eq:xbard} \\
		x\sbar(x) & = xs(x) = A_{\sbar}x^{B_{\sbar}}(1-x)^{C_{\sbar}}(1+D_{\sbar}x).\label{eq:xbars}
	\end{align}
	\label{eq:pkuparam}
\end{subequations}

In this parametrization form, we add an extra parameter $D$ to the sea quarks and put no restrictions between the individual PDF parameters, so that it might adapt the tensions between individual partons and measurements well. Setting $s = \bar{s}$ is the usual way used by many PDFs working groups because of the limited data constraints on the strange quark distribution. By taking difference of  $\sigma(W^+\bar{c})$ and $\sigma(W^-c)$, it is possible to extract information on $x(s-\bar{s})$. Recently, the newly extracted PDFs MSHT20~\cite{Bailey:2020ooq} and the reference~\cite{Faura:2020oom} obtain very small, order of $10^{-3}$, $x(s - \bar{s})$ distribution by analyzing both the dimuon and LHC data. Given some limitations in data on the strange quark distribution, we use symmetric strangeness throughout in our analysis. Actually, it was predicted~\cite{Brodsky:1996hc} that the momentum and helicity distributions of the strange and antistrange quarks are different from each other. Hopefully, this problem will be solved by analyzing relevant data from the near-future high-luminosity and high-energy experiments and with more precise theoretical calculations.

The denomination for each round of fit goes as follows:
\begin{itemize}\setlength\itemsep{1em}
	\item 	\fitone.---Applying the parametrization form of Eq.~(\ref{eq:atasparam}) with  {\hera}~\cite{Abramowicz:2015mha} and ATLAS $W/Z$ data~\cite{Aaboud:2016btc}.
	
	\item 	{\fittwo}.---Applying the parametrization form of Eq.~(\ref{eq:cmsparam}) with {\hera}~\cite{Abramowicz:2015mha} and CMS $W$ + charm and $W$-lepton charge asymmetry  data~\cite{Chatrchyan:2013mza,Khachatryan:2016pev,Chatrchyan:2013uja,Sirunyan:2018hde}.
	
	\item {\fitonep}.--- Adding ATLAS $W$ + charm data~\cite{Aad:2014xca} to {\fitone}.
	
	\item {\pkushape}.---Applying Eq.~(\ref{eq:pkuparam}) with all the data.
\end{itemize}
To get a data-driven characteristics of the PDFs, we get rid of all the restrictions set on the fit parameters, $A_{\ubar}$ = $A_{\dbar}$ and $B_{\ubar}$ = $B_{\dbar}$ = $B_{\sbar}$ in the above fits. The initial factorization and renormalization scale is still 1.9~$\mathrm{GeV^2}$. The heavy quark masses are set $m_c$ = 1.43~GeV and $m_b$ = 4.5~GeV. The strong coupling constant $\alpha_s$ is set to be $0.111$ suggested in the {\herapdf}~\cite{Abramowicz:2015mha} through a $\chi^2$ scan on $\alpha_s$. We did conduct a few rounds of fits with $\alpha_s$ values equal to 0.110, 0.111, 0.113, 0.115, 0.117, 0.118 using Eq.~(\ref{eq:atasparam}), imposing specific cuts on $Q^2 \gtrsim$ 3.5, 7.5, 10~$\mathrm{GeV^2}$, and found that fit results are almost independent on the strong coupling constant. To use as much as data, we set minimum $Q_{\mathrm{min}}^2$ = 3.5~$\mathrm{GeV^2}$ as a universal cut on the squared 4- momentum transferred. The Quality of the fit is judged by $\chi^2$/DOF (degrees of freedom (DOF), defined as the difference of the number of data points used between the number of free fit parameters), where $\chi^2$ is constructed by including the measurement, corresponding theoretical calculation, and the measurement uncertainties (statistical, correlated and uncorrelated systematic). The detailed form of the $\chi^2$ function is constructed as
\begin{equation}
\begin{split}
	\chi^2(M,\lambda) &= \sum_{i=1}^{N_{\mathrm{data}}} \frac{1}{\Delta_i^2}\left( M_i +\sum_{\alpha=1}^{N_{\mathrm{sys.corr}}}\Gamma_{\alpha i}T_i\lambda_{\alpha}-T_i\right)^2 +
	\sum_{\alpha=1}^{N_{\mathrm{sys.corr}}}\lambda_{\alpha}  + \\
	&\sum_{i=1}^{N_{\mathrm{data}}}\mathrm{ln}\frac{\Delta_i^2}{M_i^2(\delta_{i,\mathrm{stat}}^2 + \delta_{i,\mathrm{uncor}}^2)},
\end{split}
\end{equation}
where $M_i$ and $T_i$ stand for the measured and theoretical values of the cross sections. $\Gamma_{\alpha i}$, together with the nuisance parameter $\lambda_{\alpha}$, quantifies the contribution of each correlated systematic error source $\alpha$. $\delta_{i,\mathrm{stat}}$ and $\delta_{i,\mathrm{uncor}}$ are the corresponding relative statistical and uncorrelated systematical errors which are proportional to measured value. The index $i$ runs over all $N_{\mathrm{data}}$ data points while index $\alpha$ runs all of the correlated error sources. Because of the multiplicative nature of both the statistical and uncorrelated systematical errors, $\Delta_i^2$ is defined as
\begin{equation}
	\Delta_i^2 = \delta_{i,\mathrm{stat}}^2T_iM_i+\delta_{i,\mathrm{uncor}}^2M_i^2,
\end{equation}
which is corrected for statistical fluctuations in data by scaling with predicted values according to the recipe in Ref.~\cite{Aaron:2009aa}. This form of $\Delta_i^2$ leads to a logarithmic term arising from likelihood transition of $\chi^2$.
The minimization of $\chi^2$ function against the fit parameters and the nuisance parameter is done using the package {\minuit}~\cite{James:1975dr}, which is interfaced with {\xfitter}. The PDFs uncertainties arising from the uncertainties in the measurements are estimated using Hessian method~\cite{Pumplin:2001ct} adopting the same tolerance $\Delta \chi^2$ = 1 as the HERA, ATLAS and CMS groups, corresponding to 68\% confidence level (CL).

\section{Results}

Table~\ref{tab:x2} lists the corresponding $\chi^2$/DOF values for four fits showing the datasets are reasonably agreeable.
\begin{table*}[t]
	\caption{The corresponding partial, total $\chi^2$/DOF function for
the \tc{black}{four}
fits} \label{tab:x2}
	\begin{ruledtabular}
		\begin{tabular}{lllll}
			Dataset                       &{\fitone}&{\fittwo}&{\fitonep}&{\pkushape} \\
			\midrule
			HERA I+II CC $\mathrm{e^+p}$  & 49 / 39 & 49 / 39 & 47 / 39 & 44 / 39 \\
			HERA I+II CC $\mathrm{e^-p}$  & 64 / 42 & 55 / 42 & 66 / 42 & 68 /42\\
			HERA I+II NC $\mathrm{e^+p}$  & 226 / 159& 222 / 159 & 226 / 159 &  226 / 159\\
			HERA I+II NC $\mathrm{e^+p, E_p }= 820$~GeV & 66 / 70 & 70 / 70 &66 / 70&  66 / 70\\
			HERA I+II NC $\mathrm{e^+p, E_p } = 920$~GeV& 451 / 377& 449 / 377 & 450 / 377& 441 / 377\\
			HERA I+II NC $\mathrm{e^+p, E_p}  =  460$~GeV & 220 / 204& 219 / 204 & 221 / 204& 220 / 204 \\
			HERA I+II NC $\mathrm{e^+p, E_p} =  575$~GeV & 222 / 254& 224 / 254 & 222 / 254 &220 / 254\\
			\midrule
			CMS W-$\mu$ charge asymmetry $\sqrt{s}= 7$~TeV&   & 14 / 11 &  & 13 / 11\\
			CMS W-$\mu$ charge asymmetry $\sqrt{s} =8$~TeV &   & 3.2 / 11 &  & 5.5 / 11\\
			CMS W + charm $\sqrt{s} = 7$~TeV &  & 2.6 / 5 & & 1.5 /5 \\
			CMS W + charm $\sqrt{s} = 13$~TeV& & 2.5 / 5 &  &1.2 / 5\\
			\midrule		
			ATLAS $Z/\gamma^*$, $46 < m_{\mathrm{z}} < 66$ GeV  & 23 / 6&   &23 / 6  & 24 / 6 \\
			ATLAS central $Z/\gamma^*$, $66 < m_{\mathrm{z}} < 116$ GeV  & 14 / 12&  & 15 /12 & 14 / 12 \\
			ATLAS forward $Z/\gamma^*$, $66 <m_{\mathrm{z}} < 116$ GeV & 5.6 / 9&  &  6 / 9 & 4.7 / 9 \\
			ATLAS central $Z/\gamma^*$, $116 < m_{\mathrm{z}} 150$ GeV  & 6.9 / 6&  &   6.7 / 6 & 6.8 / 6 \\
			ATLAS forward $Z/\gamma^*$, $116 < m_{\mathrm{z}} 150$ GeV  & 4.5 / 6&  &  4.4 / 6  & 4.1 / 6 \\
			ATLAS $W^-\to l^-\bar{\nu}$ & 8.7 / 11&   &  10 / 11 & 9.3 / 11\\
			ATLAS  $W^+\to l^+\nu$  & 13 / 11& &  13 / 11  & 14 / 11\\
			\midrule
			ATLAS $W^+$ + $\cbar$-jet $\sqrt{s} = 7$~TeV &&& 7.5 / 11 & 7.1 / 11 \\
			ATLAS $W^-$ + $c$-jet $\sqrt{s} = 7$~TeV && &1.6 / 11  & 1.8 / 11\\
			\midrule
			Correlated $\chi^2$  & 117& 90 & 120 & 118 \\
			Log penalty $\chi^2$  & -14.18& -9.35 &  -14.13 & -6.14\\
			Total $\chi^2$ / DOF  & 1476 / 1189& 1391 / 1161 & 1491 / 1211 & 1503 / 1239 \\
		\end{tabular}
	\end{ruledtabular}
\end{table*}
\begin{figure*}[t]
	\begin{center}
		\includegraphics[scale=.35]{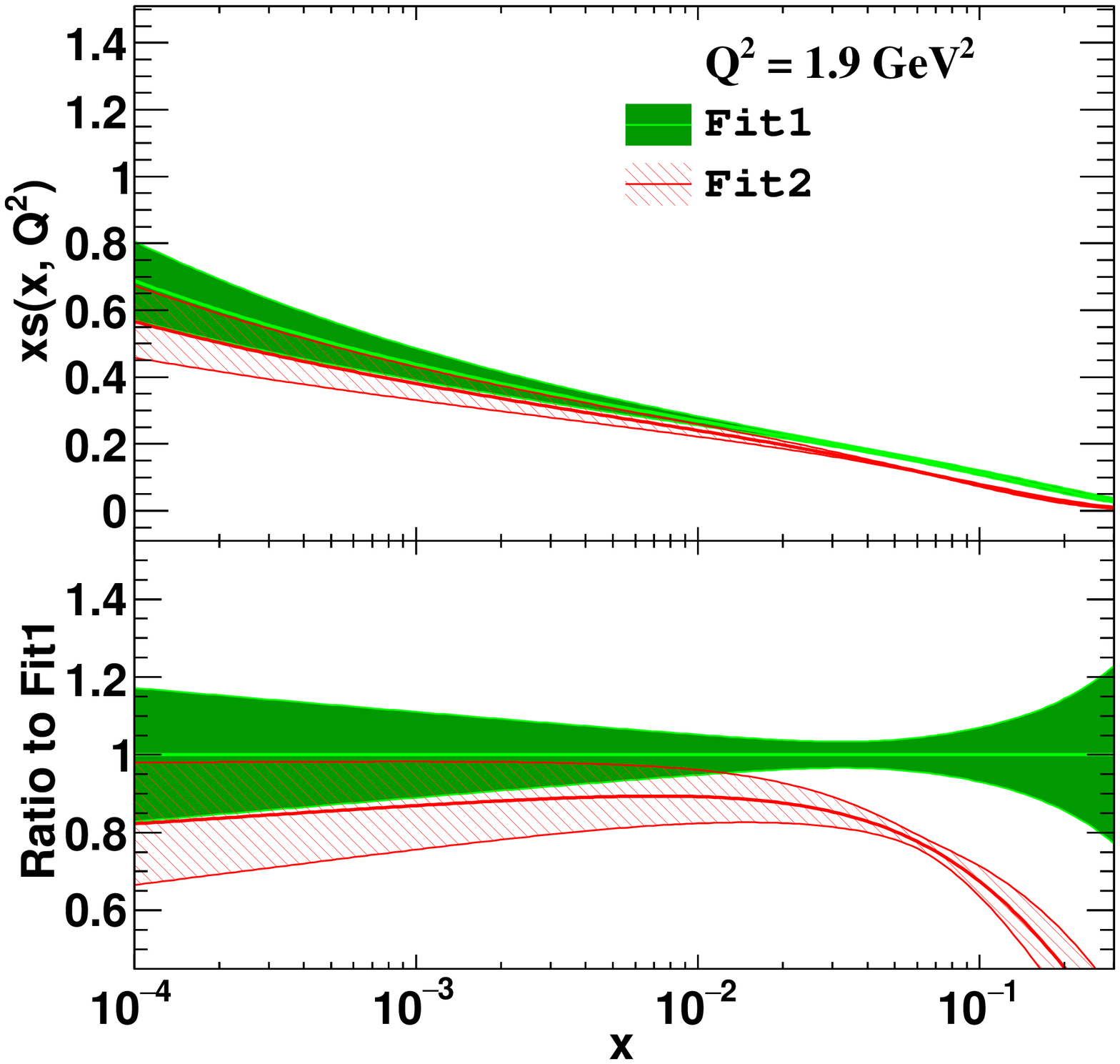}
		\includegraphics[scale=.35]{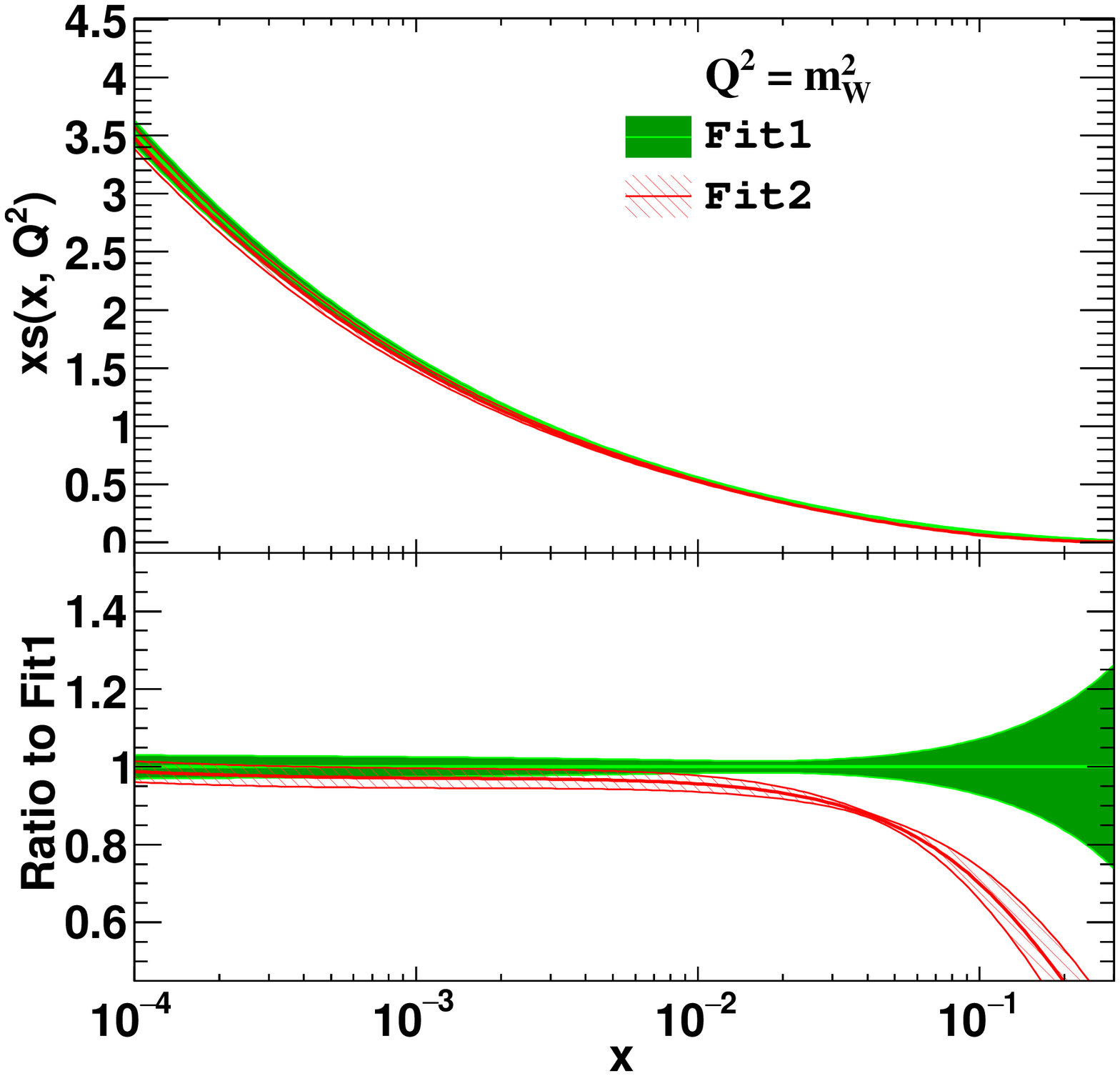}   \\
		\includegraphics[scale=.35]{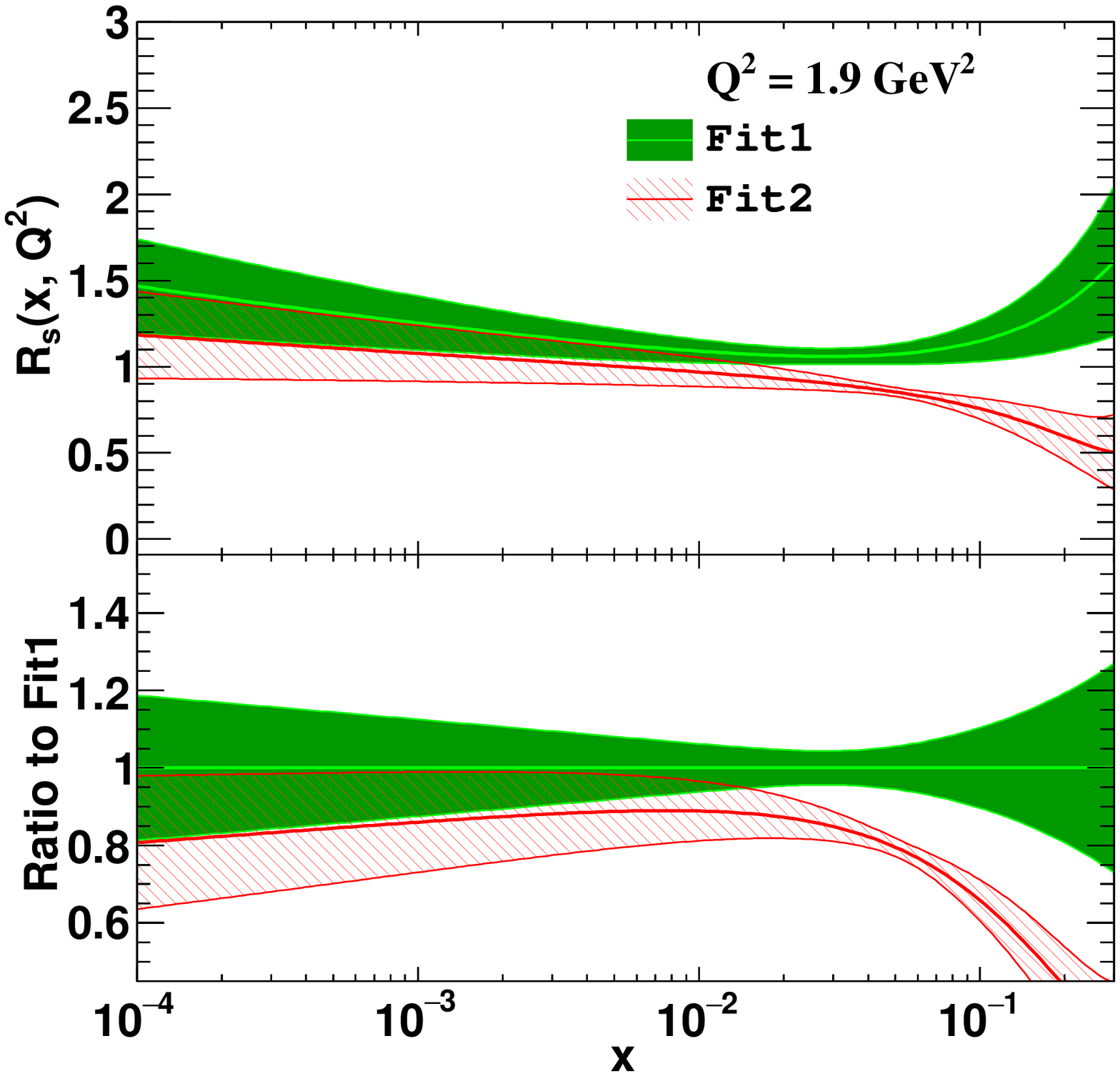}
		\includegraphics[scale=.35]{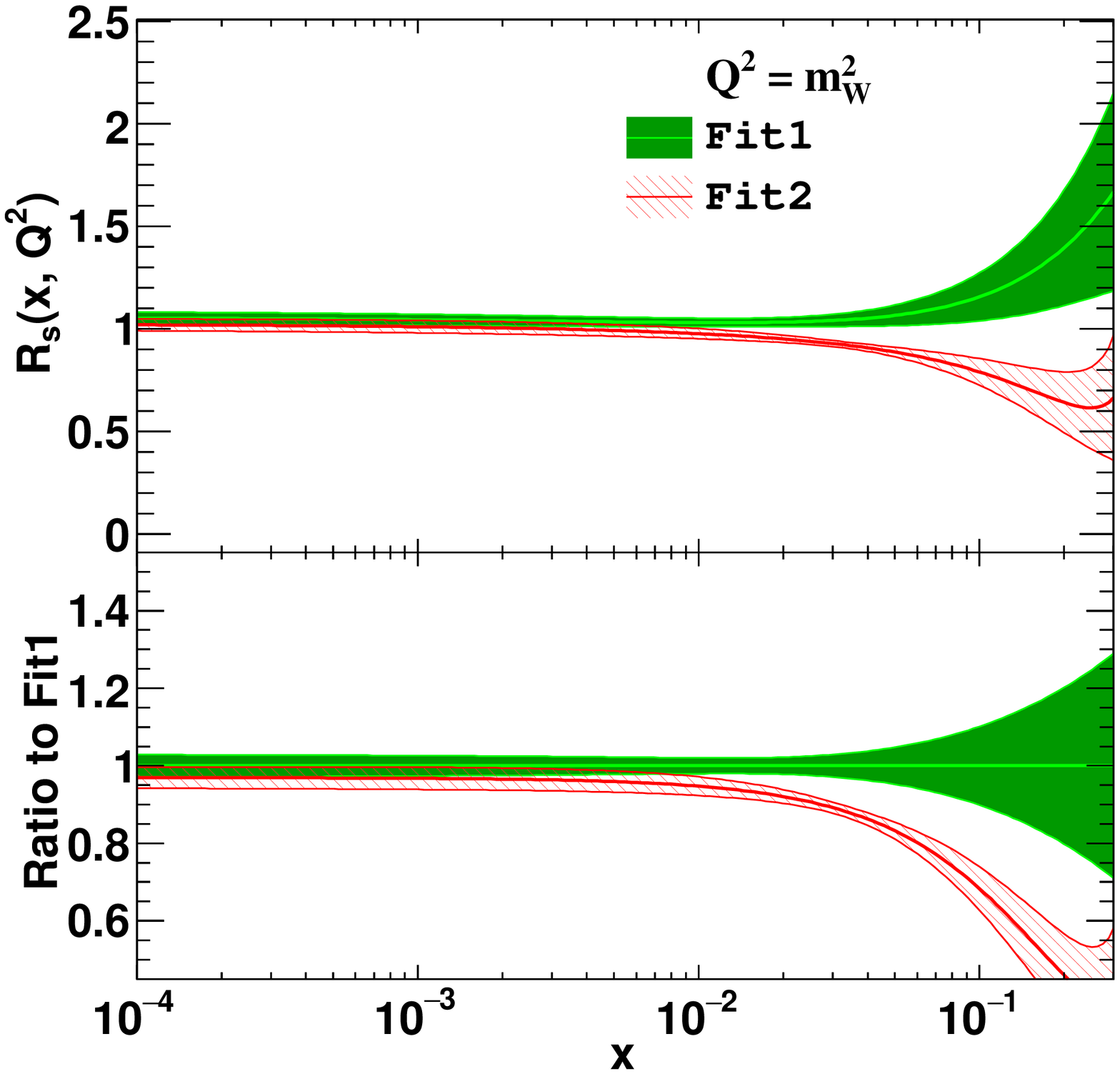}
		
		\caption{The strange PDFs and strange-to-nonstrange ratio as functions of $x$ evaluated at the factorization scale $Q^2 = 1.9~\mathrm{~GeV^2}$ and $Q^2$ = $m_{W}^2$ from {\fitone} and {\fittwo}. The upper panel shows the central distribution with error bands corresponding to 68\% CL while the bottom panel shows the ratio to {\fitone}.}
		\label{fig:2pdf_stranges}
	\end{center}
\end{figure*}

\begin{figure*}[b]
	\begin{center}
		\includegraphics[scale=.35]{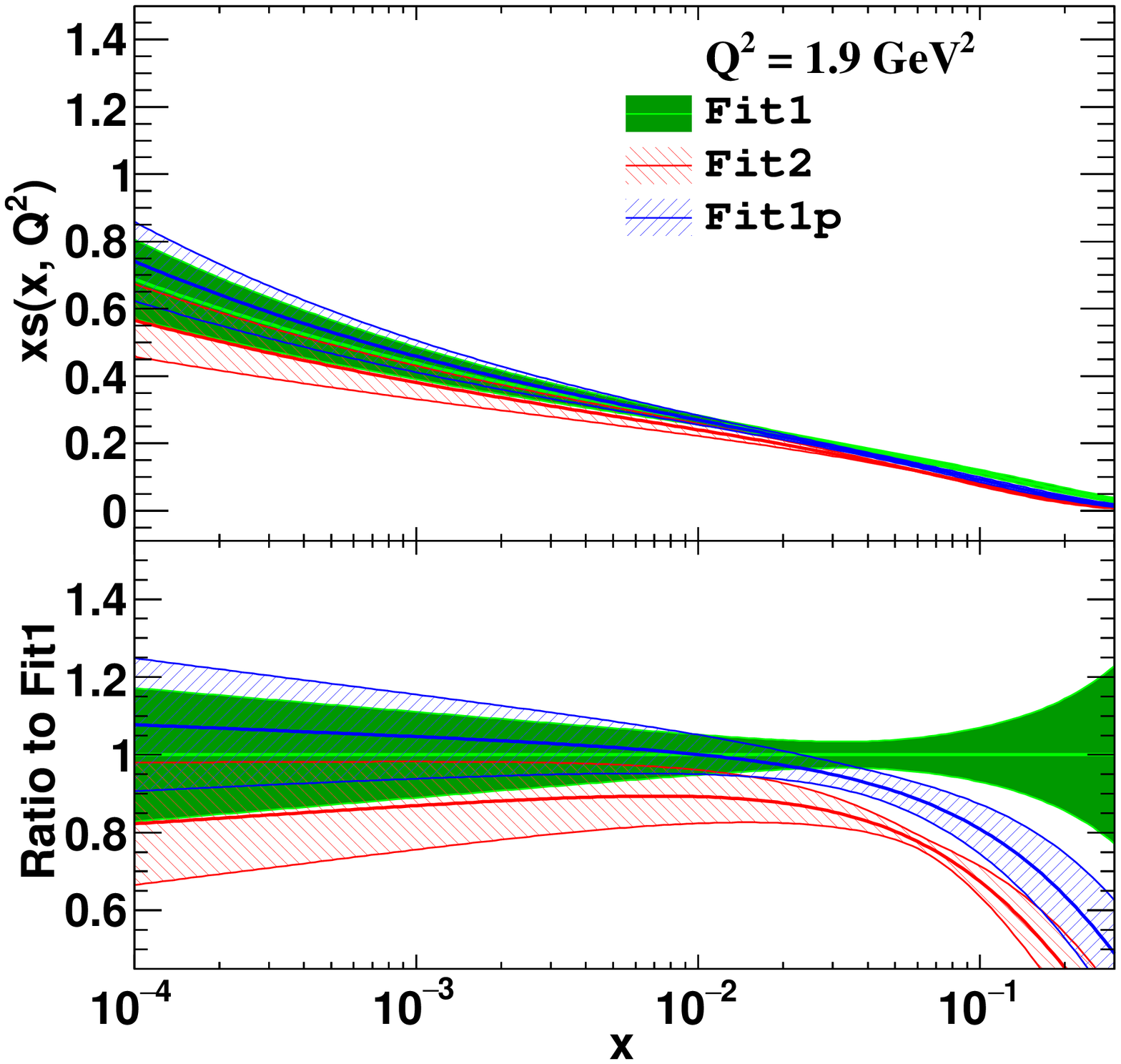}
		\includegraphics[scale=.35]{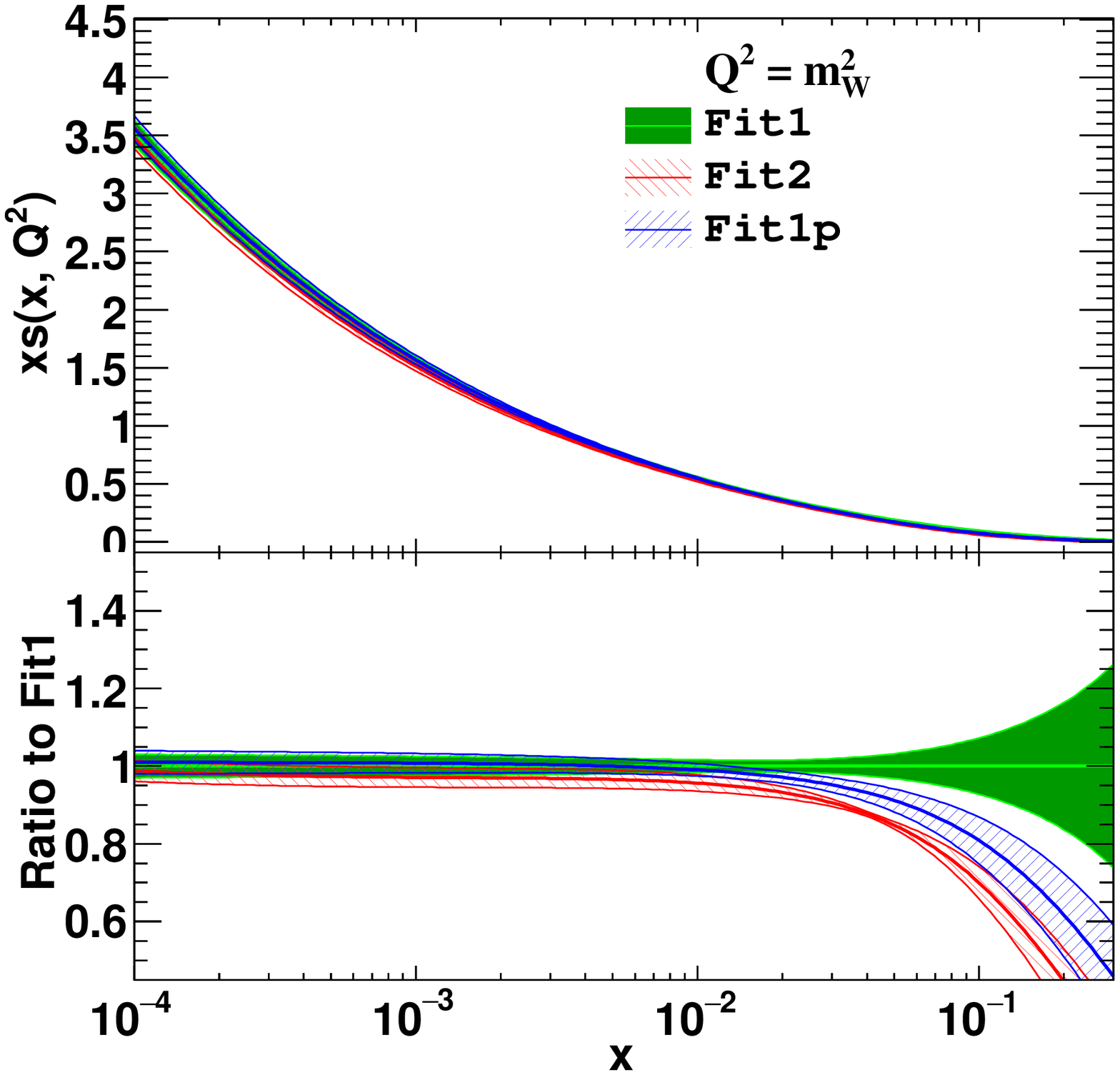}
		\includegraphics[scale=.35]{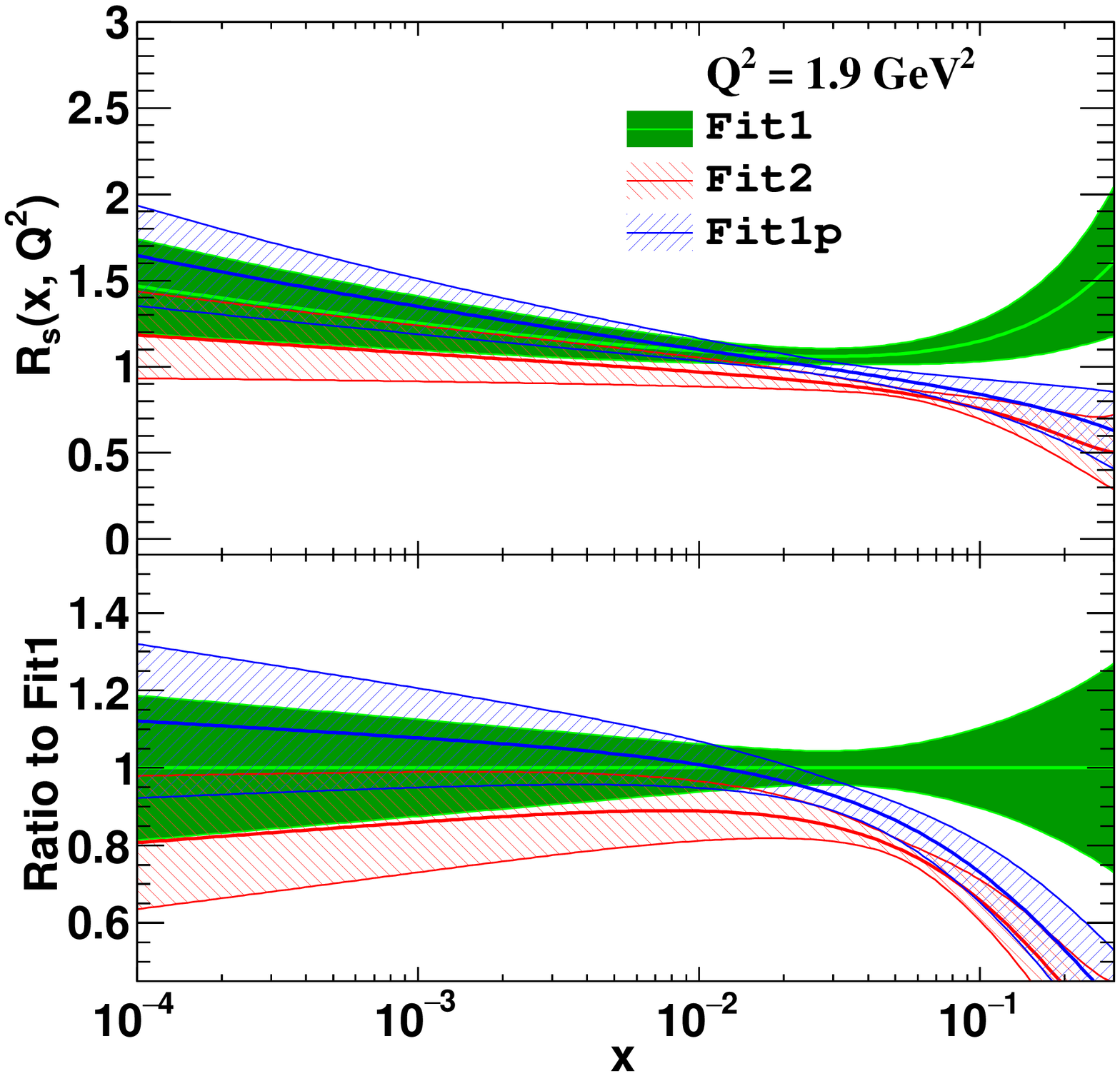}
		\includegraphics[scale=.35]{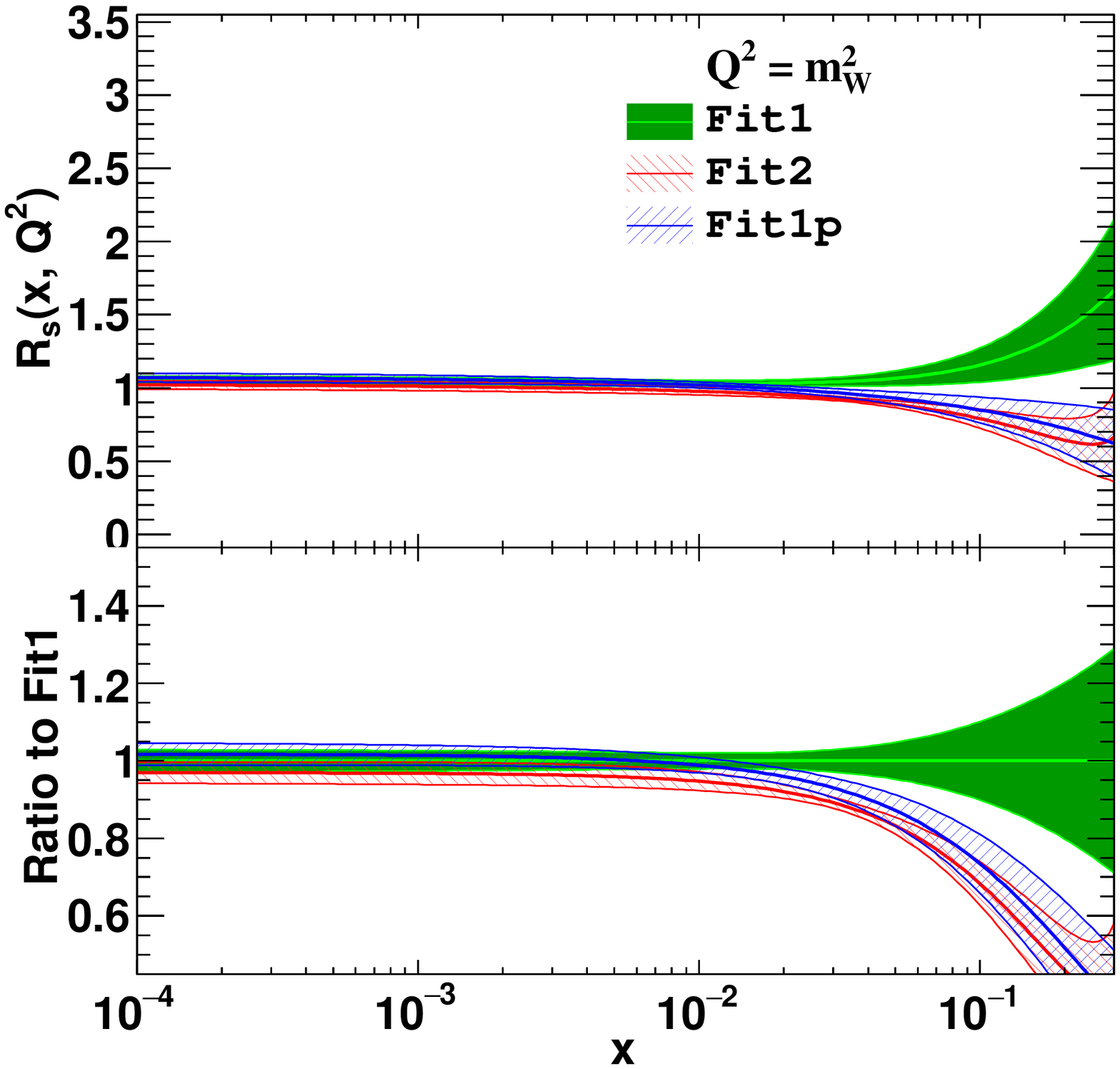}
		\caption{The strange PDFs  and strange-to-nonstrange ratio as functions of $x$ evaluated at the factorization scale $Q^2 = 1.9~\mathrm{~GeV^2}$ and $Q^2$ = $m_{W}^2$ from {\fitonep} (blue hatched band) compared to {\fitone} and {\fittwo}. The upper panel shows the central distribution with error bands at 68\% CL while the bottom panel shows the ratio to {\fitone}.}
		\label{fig:3pdf_stranges}
	\end{center}
\end{figure*}

\begin{figure*}[t]
	\begin{center}
		\includegraphics[scale=.35]{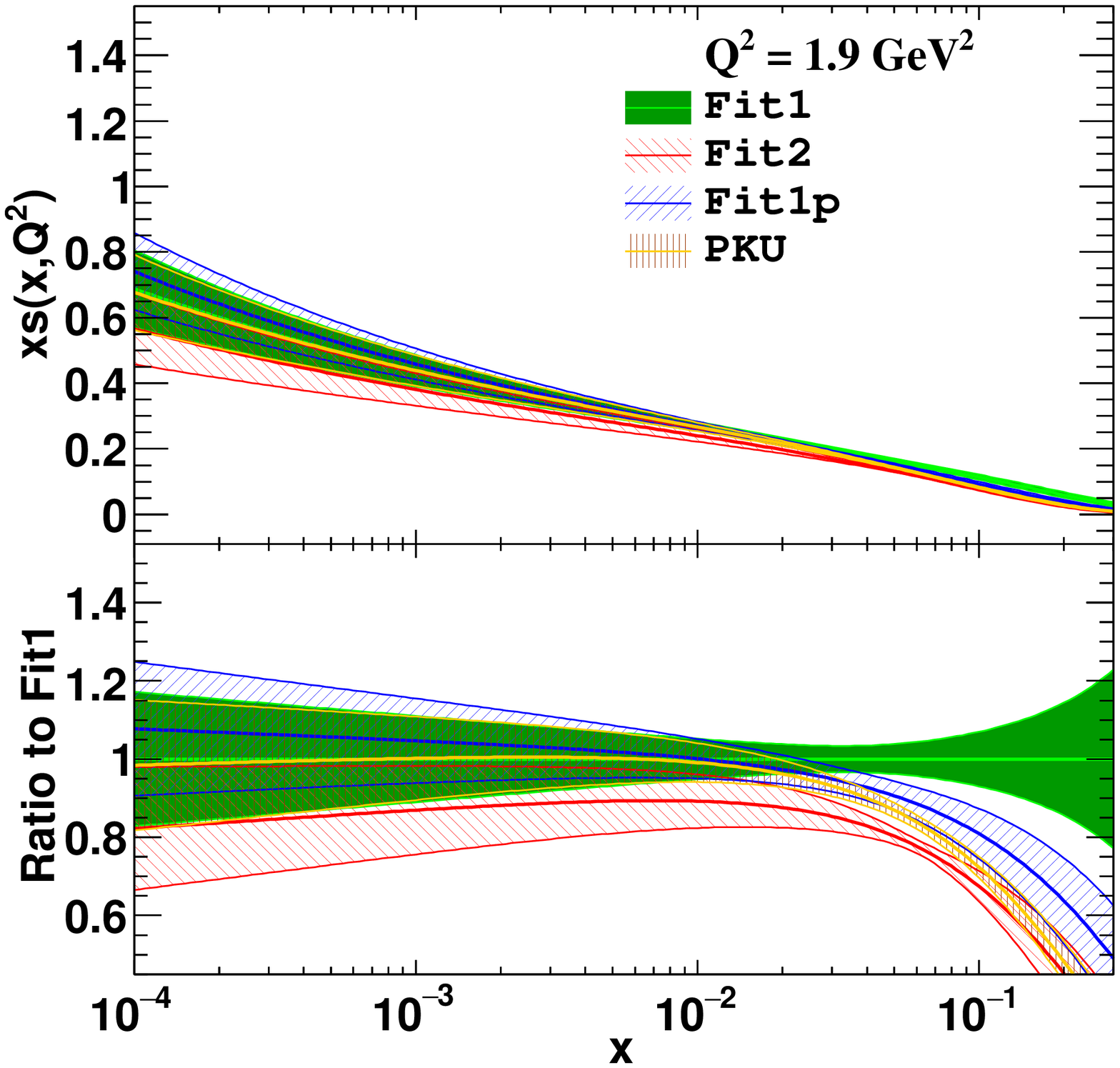}
		\includegraphics[scale=.35]{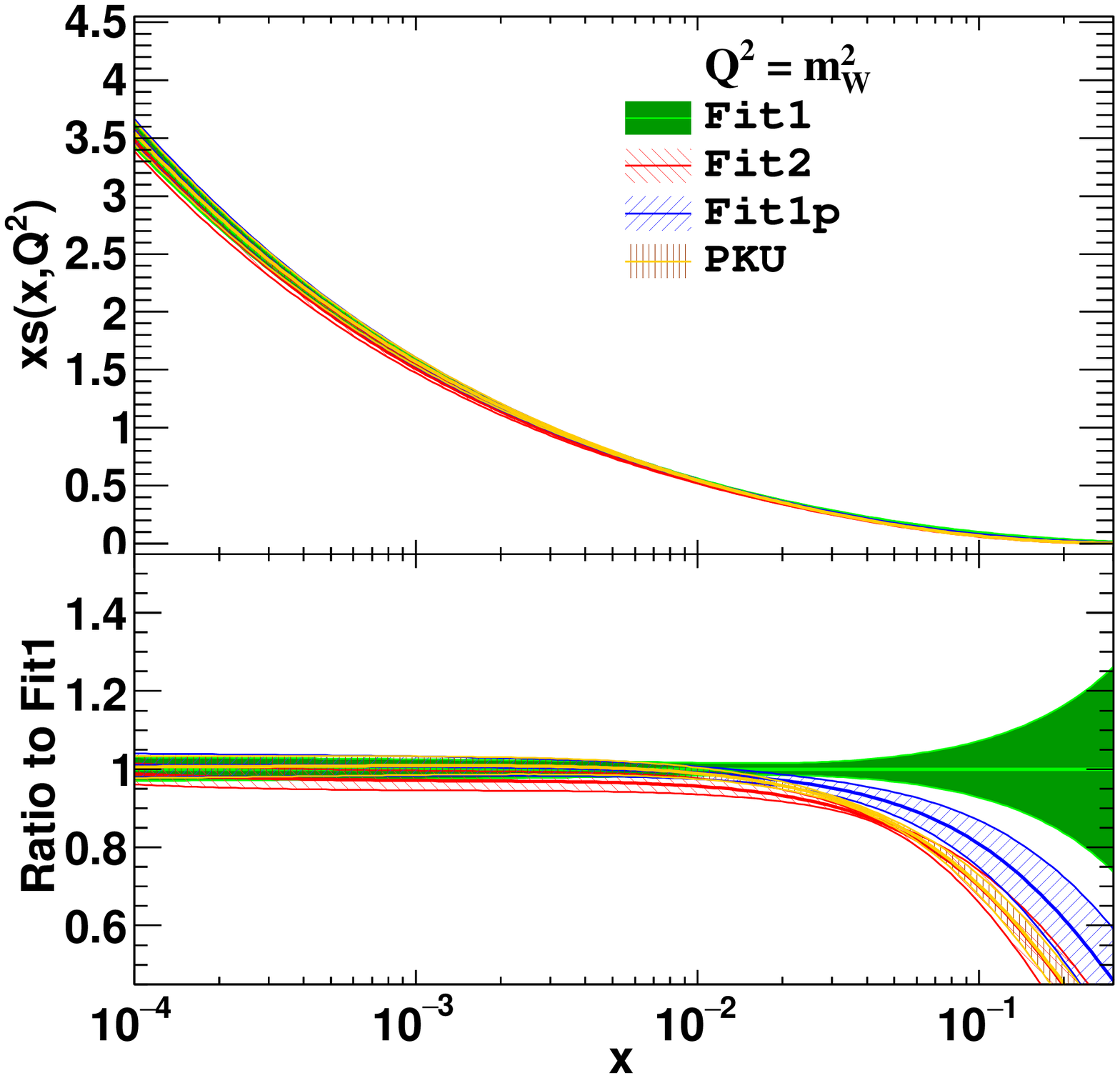} \\
		\includegraphics[scale=.35]{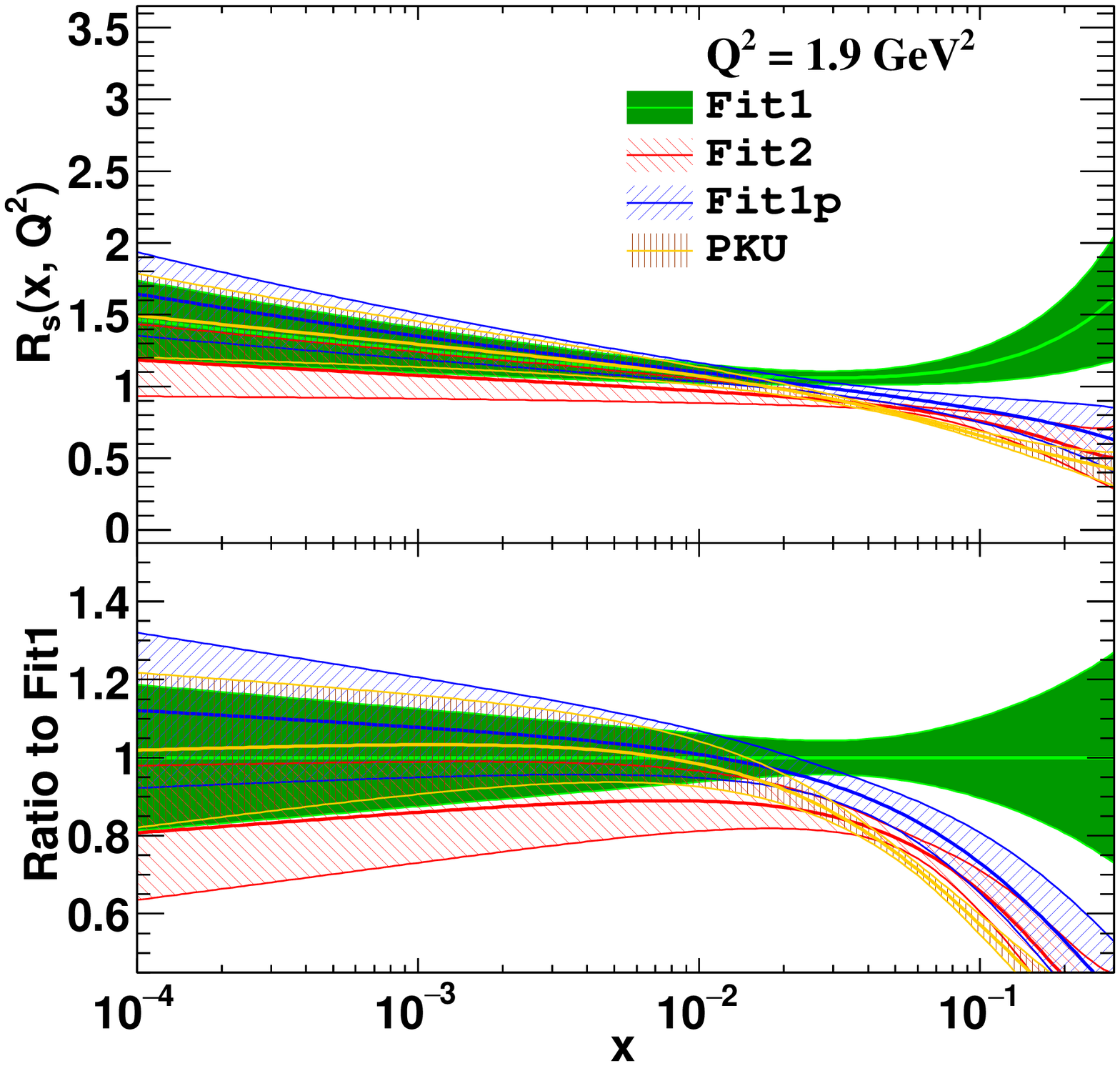}
		\includegraphics[scale=.35]{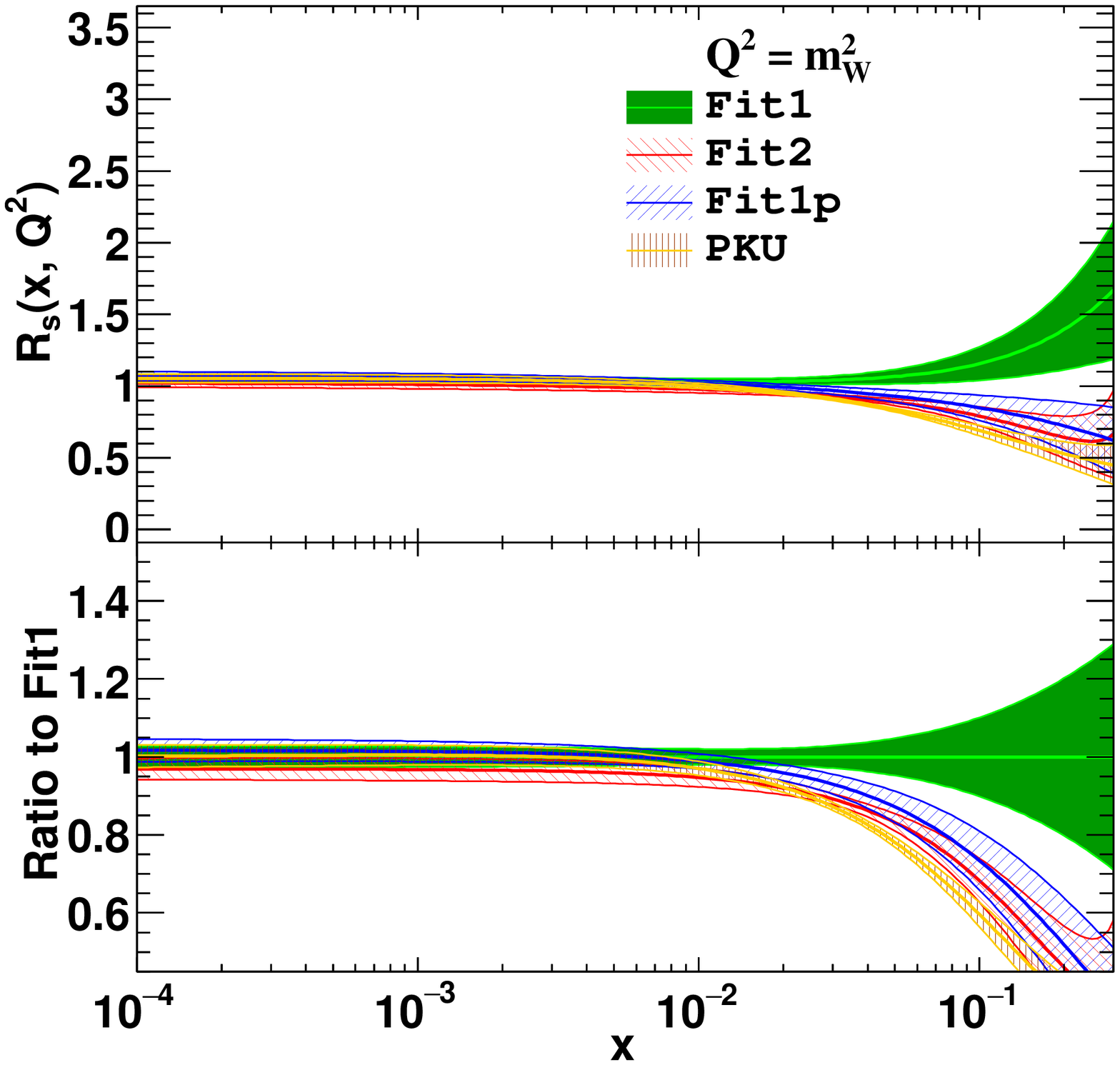}
		
		\caption{The $xs(x,Q^2)$ and $R_s$ distributions evaluated at $Q^2$ = 4 $\mathrm{GeV}^2$ as functions of $x$ for the {\pkushape} in comparison with the previous results. The bands represent 68\% CL of experimental uncertainties.}
		\label{fig:4pdf_stranges}
	\end{center}
\end{figure*}

\begin{figure*}[htbp!]
	\centering
	\includegraphics[scale=0.3]{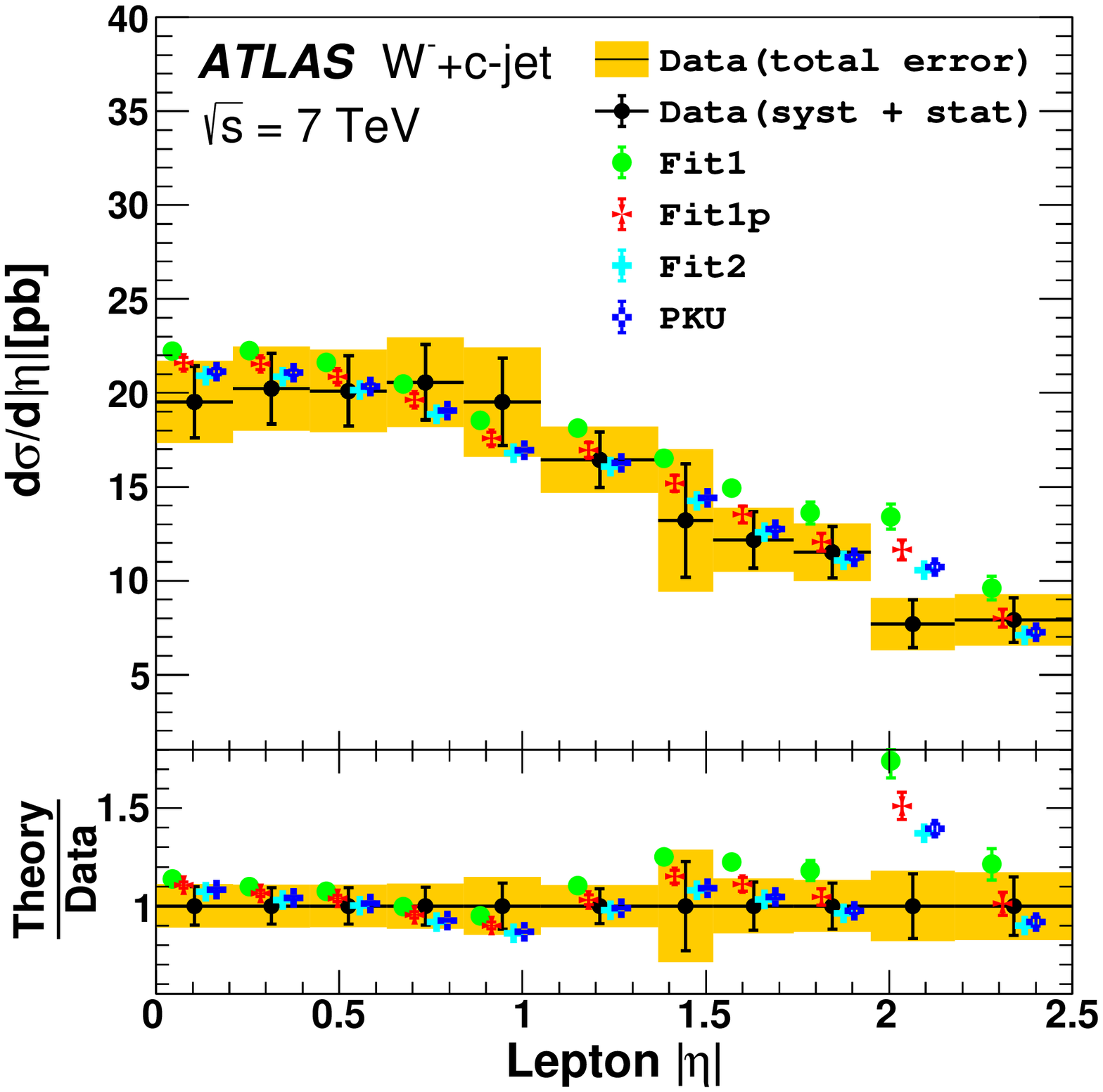}
	\includegraphics[scale=0.3]{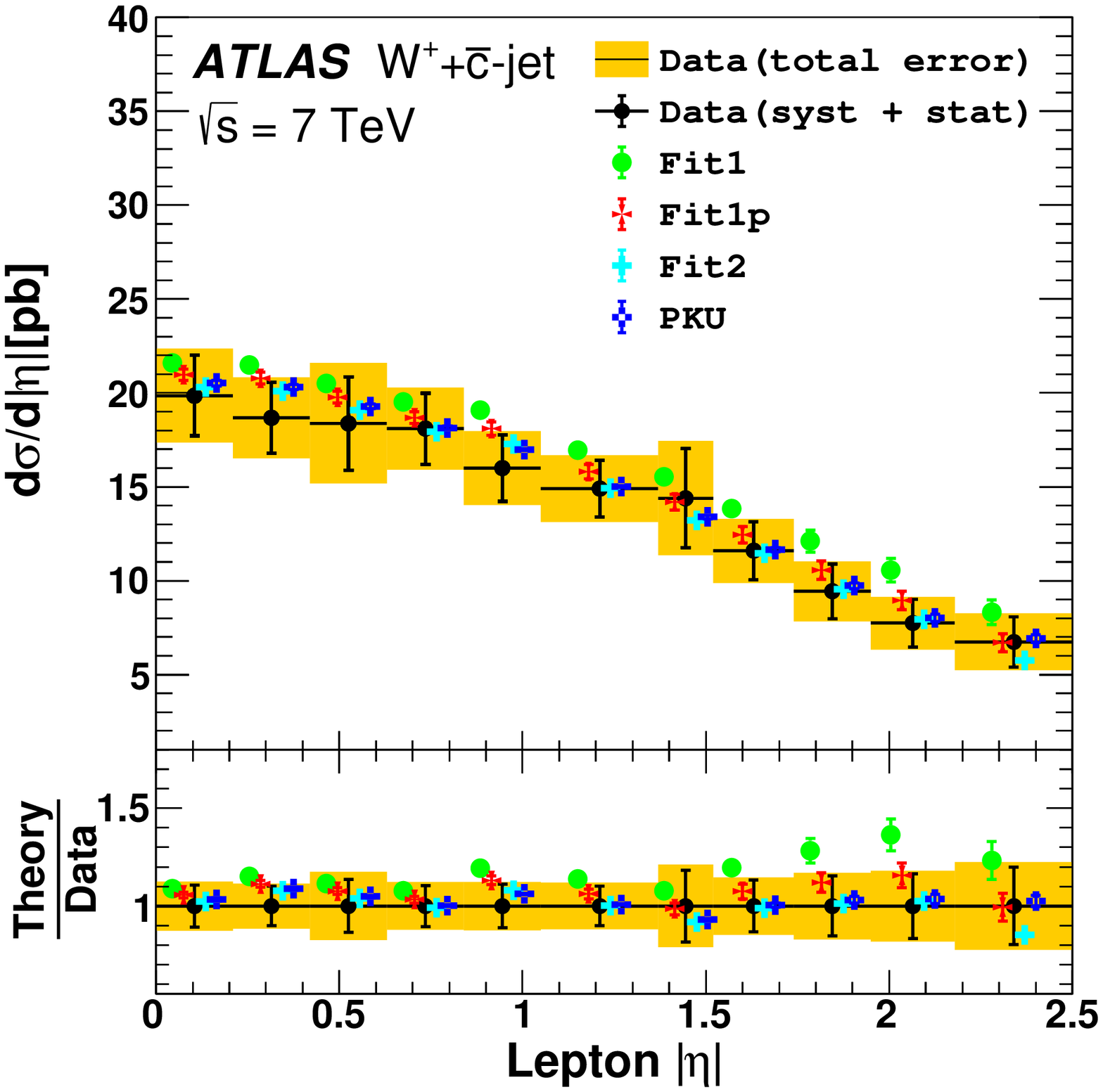}
	\caption{The measured differential cross section of ATLAS $W$ + charm  production as a function of lepton $|\eta|$  compared to predictions computed with the resulted PDF sets: $W^{-}$+c-jet (left panel) and $W^{+}$+$\bar c$-jet (right panel). The measurements are shown with black error bars for systematic + statistical uncertainty and with filled bands for total uncertainty (including uncorrelated certainty). The predictions are obtained with PDF uncertainties at 68\%CL.}
	\label{fig:atsWC}
\end{figure*}

\begin{figure*}[hbt!]
	\centering
	\includegraphics[scale=0.3]{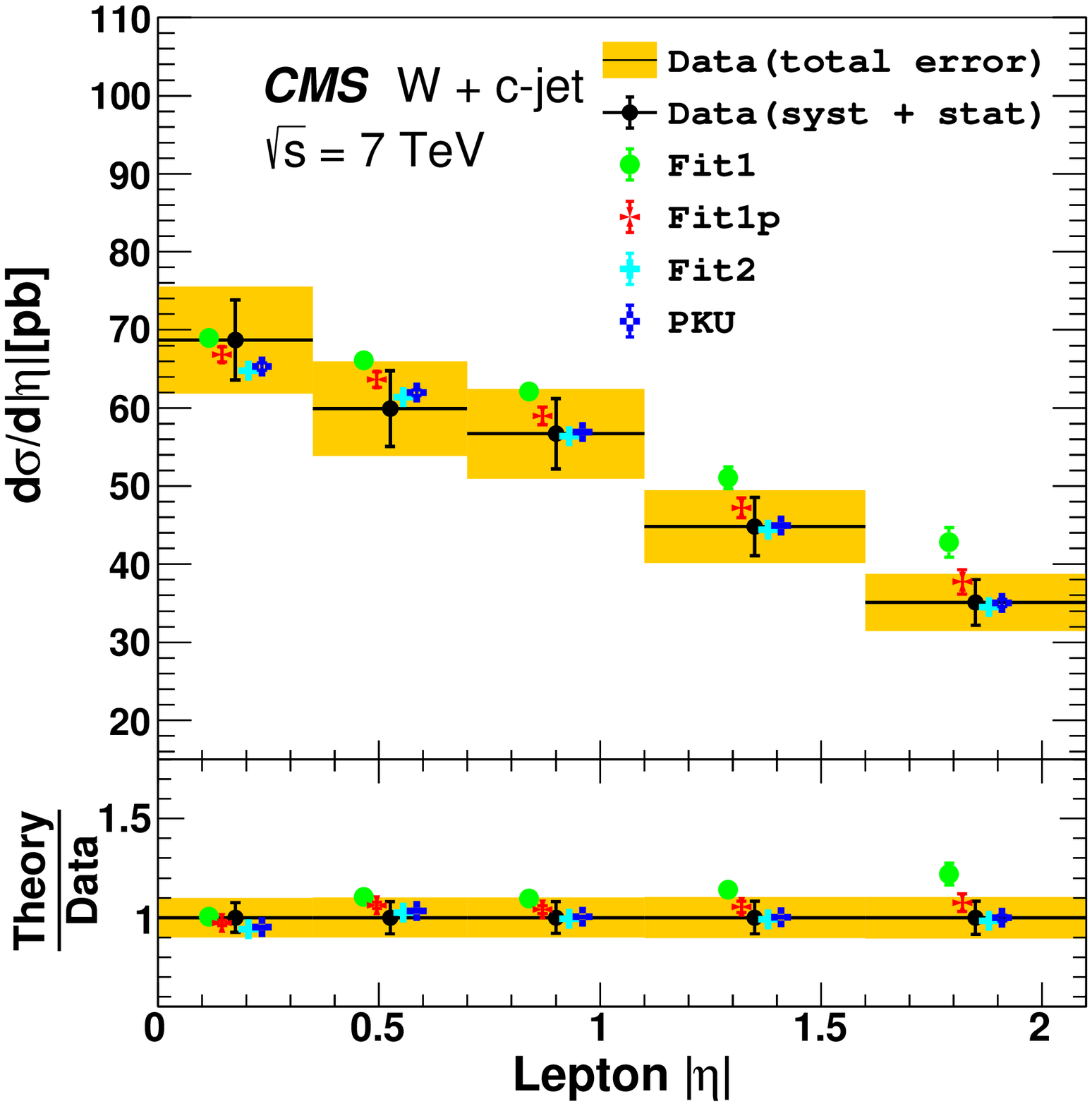}
	\includegraphics[scale=0.3]{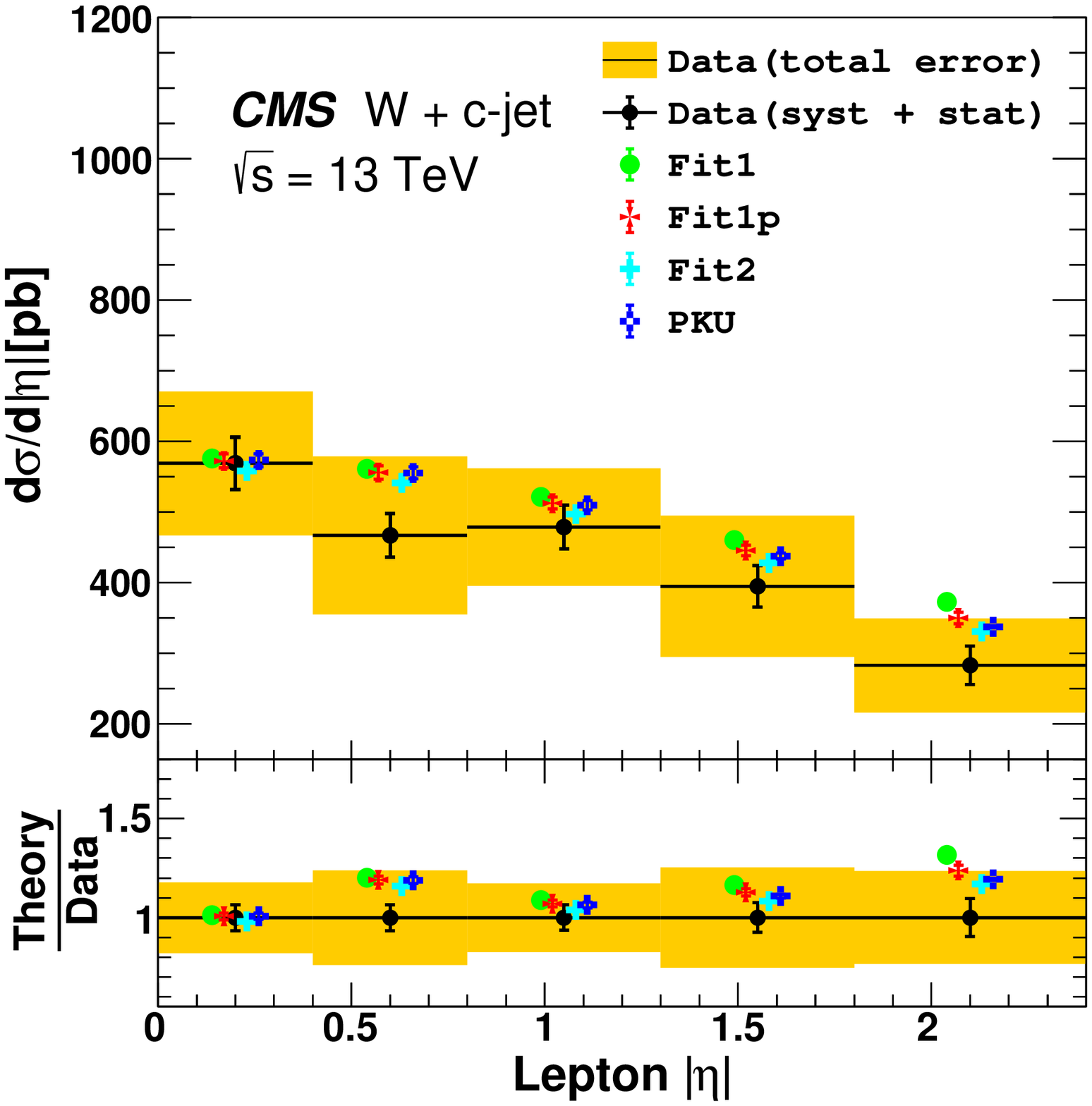}
	\caption{The measured differential cross section of CMS $W$ + charm  production as a function of lepton $|\eta|$  compared to predictions computed with the resulted PDF sets:$\sqrt{s} = $ 7~TeV (left panel) and $\sqrt{s} = $13~TeV (right panel). The measurements are shown with black error bars for systematic + statistical uncertainty and with filled bands for total uncertainty (including correlated uncertainty). The predictions are obtained with PDF uncertainties at 68\%CL.}
	\label{fig:cmsWC}
\end{figure*}

In the first place, we show the resulted strange quark PDFs and the ratio $R_s$, with ratio to {\fitone}, extracted from ATLAS $W/Z$ and CMS $W$ + charm data in Fig.~\ref{fig:2pdf_stranges} applying Eq.~(\ref{eq:atasparam}) ({\fitone}) and Eq.~(\ref{eq:cmsparam}) ({\fittwo}) at the factorization scale $Q^2$= 1.9~$\mathrm{GeV^2}$ and the scale at which the $W$ boson is produced. The $xs(x,Q^2)$ distributions display overall agreement within the uncertainties at both scales. However, it is clear to see that they differ distinctly from each other toward large $x$. Similarly, the $R_s$ distributions on the bottom pictures of Fig.~\ref{fig:2pdf_stranges} follow the same trend that the strange quark PDF has, thus indicating that the ATLAS $W/Z$ and CMS $W$ + charm data have some tensions at large momentum fraction. The central distribution of strangeness from the fit of ATLAS $W/Z$ data, {\fitone}, whether in terms of parton level or the ratio $R_s$, is slightly higher than the one obtained in {\fittwo}. The ratio $R_s$ indeed has a significant $x$-dependent property that coincides with the results obtained by CMS analysis of $W$ + charm data~\cite{Sirunyan:2018hde}.
The incompatibility between strange quarks is around 3$\sigma$.

So, the reason that gives rise to the different interpretations of strangeness in terms of the ratio $R_s$ between the original ATLAS and CMS analyses lies in the different data structure or physics processes, and, possibly, the somewhat different parametrization forms with flexible parameters of PDFs also play some role; if the sea quarks are related with each other by setting $A_{\ubar}$ = $A_{\dbar}$ and $B_{\ubar}$ = $B_{\dbar}$ = $B_{\sbar}$, they shape out the ATLAS-type $R_s$ as the ATLAS Collaboration obtained in~\cite{Aaboud:2016btc,ATLAS:2021qnl}. On the other hand, they produce the similar CMS-type of $R_s$ if the parameters are set free.
As for the disparity on this ratio arising toward large momentum fraction in {\fitone} and {\fittwo}, this is mainly because {\fitone} does not include the $W$ + charm production data that have more significance than $W/Z$ process when it comes to the strange quark PDFs.

In the second place, the resulted strange quark PDFs and the ratio $R_s$ of {\fitonep} are depicted in Fig~\ref{fig:3pdf_stranges} in comparison with {\fitone} and {\fittwo}. After the inclusion of ATLAS $W$ + charm data, $xs(x,Q^2)$ distribution of {\fitonep}, depicted with blue hatched band, demonstrates a decent agreement with {\fitone} and {\fittwo} at $x \lesssim 10^{-2}$, while being compatible with $xs(x,Q^2)$ of {\fittwo} toward large $x$. The $R_s$ of {\fitonep}, shown on the bottom, also manifests compatibility with that of {\fittwo} with a sharp deviation from that of {\fitone} at $x \gtrsim 10^{-2}$. This compatibility is highly because that the $W$ + charm data exert more constraints on the strange quark PDFs in {\fitone}.

Based on these results, we can safely state that the observed disparity in the interpretations of strangeness by the ATLAS and CMS groups, can be ascribed mainly to the different usage of data from the different physical processes. The outcome would be consistent if this factor were handled properly \tc{black}{with more flexible parametrization forms}.

The combined usage of the inclusive $W/Z$ measurements with the $W$ + charm data and $W$-lepton charge asymmetry provides a better sensitivity on the strangeness than used separately. Having clarified the discrepancy and its origin between ATLAS and CMS interpretations of the strange quark PDFs, a fairly exact distribution of the strange quark PDFs obtained through the {\pkushape} 
is shown in Fig.~\ref{fig:4pdf_stranges} in comparison with {\fitone}, {\fittwo}, and {\fitonep} at the initial and $W$ mass scale. It is clear that the strange quark PDFs of {\pkushape} are still in accordance with that of {\fittwo} and {\fitonep} throughout entire $x$ while agreeing with {\fitone} only at $x \lesssim 10^{-2}$ and deviating from it toward large $x$. As {\fitonep} and {\fittwo} display, {\pkushape}
also manifests a smoothly $x$-dependent behavior of $R_s$ at $Q^2 = 1.9~{\mathrm{GeV^2}}$, which is declining faster than other fits toward large $x$. The ratio $R_s$ distribution at $Q^2$ = $m_W^2$ tends to unity at $x \lesssim 0.02$, suggesting $(s + \bar s)  = (\bar u + \bar d) $. Generally, we can confirm that the strange quark PDFs, from both the ATLAS $W/Z$ and CMS $W$ + charm data, are fairly consistent within uncertainties in smaller momentum fraction regions and manifest a tension toward large momentum fraction. This tension between fades away when the same physical processes are used  with flexible parametrization forms. The strange quark density of the proton is only suppressed at $x > 0.023$ and enhanced toward smaller $x$.

As confirmed, the inclusion of $W$ + charm data together with $W/Z$ from both the ATLAS and CMS experiments produce well-constrained strangeness. We calculate the cross section of $W$ + charm production of both the ATLAS and CMS experiments with the extracted PDF sets to see how much the strange quark PDFs from {\pkushape} improve the theoretical results. This is pictured in Figs.~\ref{fig:atsWC} and~\ref{fig:cmsWC} compared with the measurements. The predictions of both the ATLAS and CMS $W$ + charm measurements agree with the measurements quite well. However, it is worth noting that the calculation obtained from {\fitone} is apparently bigger than the measurements at $|\eta|\approx 2$. The deviations at the large pseudorapidity have a relation with strange quark density at the corresponding momentum fraction $x$. This means that the deviation appears when there is an increase in the strange quark density. Understandably, it is the large-$x$ distribution of the strange quark that leads to the deviation. This tells us that the uncertainty of the strange quark density in {\fitone} by fitting only the ATLAS $W/Z$ data is large at large momentum fraction.

\section{Conclusions}

The shapes of the strange quark PDFs of the proton extracted from LHC data were interpreted in different ways: the ATLAS statement of fully enhanced strangeness and the CMS statement of suppressed one which supports the same idea obtained from fixed target neutrino-nuclear collision experiments. In this paper, through four rounds of analyses of the {\hera}, LHC $W/Z$,  $W$ + charm production, and $W$-lepton charge asymmetry data, \tc{black}{i.e.,} {\fitone}, {\fittwo}, {\fitonep} and {\pkushape}, we have investigated the exact distribution of the strange quark and its ratio to nonstrange sea quarks implementing the ATLAS and CMS analyses set up with free sea quark parameters using {\xfitter} program, and resolved tensions between ATLAS and CMS data, and finally extracted fairly well-shaped strange quark densities. At first,
both the ATLAS $W/Z$ and CMS $W$ + charm data are proved, through {\fitone} and {\fittwo} fits, to be partially consistent with the relaxed sea quark parameters in terms of strange quark and its ratio to nonstrange sea quark distributions at $x \lesssim 10^{-2}$. The inconsistency toward large $x$ between strange quark density and the ratio $R_s$  from the ATLAS and CMS Collaborations suggests that there is tension between ATLAS 7$~\mathrm{TeV}$ $W/Z$ and CMS 7 and 13$~\mathrm{TeV}$ $W$ + charm data. It is shown in {\fitonep} that this tension can be eliminated by fitting both the $W/Z$ and $W$ + charm data simultaneously. The shapes of the strange quark PDFs and the ratio obtained in our analysis are purely data driven. There is no handmade constraint on the fit parameters. The $W$ + charm cross section is calculated with the obtained PDFs and deviations from measurement at the large lepton $|\eta|$ are observed.
The PDFs obtained from only the {\hera} and ATLAS $W/Z$ measurement produces larger cross section values than the $W$ + charm data at the large lepton $|\eta|$, suggesting a large uncertainty of the proton strangeness from fitting only the ATLAS $7~\mathrm{TeV}$ $W/Z$ data at large $x$.
When the {\hera}, ATLAS $W/Z$, CMS $W$ + charm and $W$-lepton data are used in combination, further strengthened by ATLAS $W$ + charm measurement, with a more flexible parametrization form being applied, analyzed in {\pkushape}, the resulted distribution of the strange quark density gets improved and the ratio $R_s$ shows more $x$-dependent property. The strangeness is gradually suppressed at larger $x$ but it is highly enhanced toward smaller $x$ with its magnitude becoming comparable to that of the light-flavor up and down sea quarks.

\section{Acknowledgment}
We thank the \xfitter~team for providing useful help and discussions on the usage of their awesome program. We are also so grateful to Katerina Lipka from DESY for providing the corresponding grid file of CMS $W$ +  charm production at $\sqrt{s} = 13~\mathrm{TeV}$ for the corresponding theoretical calculation. This work is supported by the National Natural Science Foundation of China (Grant No.12075003).

\end{document}